\documentclass{IEEEtaes_no_copyright_no_footer}
\usepackage{graphicx,color} 
\usepackage{xcolor,url,xurl}
\usepackage{dirtytalk}
\usepackage{hyperref}
\usepackage{wrapfig}
\usepackage{cite}

\newcommand{\ignore}[1]{}

\title{A Taxonomy of Space Infrastructures Attacks and Defenses}

\author{JOSE LUIS CASTANON REMY}
\affil{~Laboratory for Cybersecurity Dynamics, Department of Computer Science, University of Colorado Colorado Springs} 

\author{SHOUHUAI XU}
\affil{~Laboratory for Cybersecurity Dynamics, Department of Computer Science, University of Colorado Colorado Springs}

\date{}
\begin{document}
\bstctlcite{IEEEexample:BSTcontrol}






\maketitle

\begin{abstract}
Space infrastructures represent an emerging domain that is critical to the global economy and society. However, this domain is vulnerable to attacks, including cyber attacks and other kinds of attacks. To enhance the resilience of this domain, we must understand these attacks that can be waged against it and the defenses that can be employed to mitigate these attacks. The status quo is that there is neither a systematic understanding of these attacks against, nor defenses for, space infrastructures 
despite their clear importance in guiding systematic analysis of space security and future research. In this paper, we fill the void by proposing the first systematic taxonomy of attacks against, and defenses for, 
space infrastructures. We hope this paper will inspire a community effort at refining the taxonomy towards a widely used one. 
\end{abstract}


\section{Introduction}

Space infrastructures-enabled 
services have been embedded into our daily life and have become essential to the worldwide economy and society. This statement can be supported by reports of the Space Foundation \cite{spacefoundation2022,spacefoundation2023}, which state that in 2021 the global space economy was approximately US\$469B and will increase to US\$800B in 2027.
However, space infrastructures are vulnerable to attacks, including cyber attacks and non-cyber attacks.  
The status quo is that there is no systematic taxonomy of attacks against, and defenses for,
space infrastructures, despite the endeavors on systematizing literature cyber attacks \cite{remy2025sok,falco2021security,ceccato2021generalized,schmidt2016survey, xiao2018secure, li2019physical, guo2021survey, tedeschi2022satellite, meng2022survey, pavur2022building, yuan2023authenticating, chen2023satellite,koisser2024orbital, manulis2021cyber}, real-world attacks against space infrastructures \cite{ear2023characterizing}, and literature 
defenses for space infrastructures \cite{pirayesh2022jamming,jia2018anti,mirkovic2004taxonomy,ferdous2023review,remy2025sok,pavur2022building}. The importance of a systematic taxonomy can never be overestimated because it can not only deepen our understanding of the problem but also guide future studies in designing defensive solutions. The lack of a systematic taxonomy motivates the present study.

\smallskip

\noindent{\bf Our Contributions}. This 
paper makes two contributions. The first contribution is a systematic taxonomy of attacks against 
space infrastructures. 
The taxonomy includes three categories of attacks: 
(i) {\em counterspace attacks}, which attempt to destroy or disrupt space infrastructures; (ii) {\em electromagnetic attacks}, which attempt to deny, disrupt, or deceive space infrastructures; and (iii) {\em cyber attacks}, which attempt to disrupt space infrastructures-enabled services, control over spacecraft, and exfiltrate data.
The attack taxonomy has two characteristics. 
(a) It accommodates a unique feature of attacks against space infrastructures: they are not only vulnerable to cyber attacks, which are common to other kinds of infrastructures,
but also vulnerable to non-cyber attacks (i.e., electromagnetic and counterspace attacks), which are symbolic to space infrastructures.
(b) It is well aligned with the way of thinking of two industrial standards: the MITRE ATT\&CK framework \cite{ATTCK}, which describes cyber attacks against terrestrial Information Technology (IT) networks; and the SPARTA framework \cite{SPARTA}, which describes 
attacks against space segments (e.g., satellites). This is because for each attack we consider: (b.1) attacker {\em objective}, or {\em tactics} in the terminology of \cite{ATTCK,SPARTA}; and (b.2) attacker {\em capabilities}, or {\em techniques} and {\em sub-techniques} in the terminology of \cite{ATTCK,SPARTA}. This description of attacker objectives and capabilities is also consistent with the notion of {\em threat models} used by academic researchers and curriculum. This alignment with industrial frameworks and academic literature is important to unify professionals' way of thinking.
However, our taxonomy goes much beyond \cite{ATTCK, SPARTA} and the existing academic literature because: (i) we accommodate counterspace, electromagnetic, and cyber attacks, while noting that counterspace attacks  are only briefly mentioned in \cite{SPARTA, falco2021security};
(ii) we categorize all attacks based on their objectives (the ``what'') and then their capabilities (the ``how''); 
and (iii) we specify the {\em entry point} and {\em impact point} of each attack in a space infrastructure. 


\ignore{
{\color{purple}Note, although
the ``impact'' tactic in \cite{ATTCK, SPARTA} can be compared to attacker objectives, and the techniques in 
\cite{ATTCK, SPARTA}
can be used to further describe attacker capabilities, by identifying attacker objectives and attacker capabilities, we can clearly describe an attack, which is not possible 
just by using tactics and techniques (i.e., pairing tactics and techniques to describe an attack requires additional steps, as demonstrated in \cite{ear2023characterizing}).} 
}




The second contribution is a taxonomy of defenses for space infrastructures. 
The taxonomy includes three categories of defenses that are respectively mapped to the three categories of attacks: (i) {\em counterspace defenses}, which aim to cope with counterspace attacks; (ii) {\em electromagnetic defenses}, which aim to cope with electromagnetic attacks; and (iii) {\em cyber defenses}, which aim to detect and react to cyber attacks. The taxonomy has two characteristics. 
(a) It specifies the {\em deployment point} of a defense mechanism in a space infrastructure. (b) It classifies defense mechanisms into two kinds: 
{\em preventive}, and {\em reactive} 
defenses, which offer a systematic way of thinking in treating defenses.

\smallskip

\noindent{\bf Related Work}.
Regarding attack taxonomy, there are four prior studies. There is a taxonomy of {\em security risks} to space infrastructures \cite{falco2021security}, including {physical}, {digital}, {organizational}, and {regulatory} risks. 
There is a taxonomy of {\em threats} against space assets \cite{spacethreatlandscape}, including 10 threats (i.e., {nefarious}, {eavesdropping}, {interception}, {jamming}, {physical}, {unintentional}, {failures}, {outages}, {disaster}, and {legal}). 
There is a systematization of cyber and electromagnetic attacks against space infrastructures reported in academic literature \cite{remy2025sok}. There is a systematization of real-world cyber attacks against space infrastructures   \cite{XuCNS2023} that focuses on addressing the problem of missing data. 
These taxonomies are different from ours because: (i) we consider counterspace, electromagnetic, and cyber attacks; (ii) we systematically describe the attacker {\em objectives} and {\em capabilities} of each attack, while aligning with ATT\&CK \cite{ATTCK} and SPARTA \cite{SPARTA} as well as the academic literature; and (iii) we describe attack {\em entry points} and {\em impact points} for each attack.

Regarding defense taxonomy, there are five most relevant prior studies.
There is a survey on anti-jamming defenses geared towards wireless networks \cite{pirayesh2022jamming}, and satellite communications \cite{jia2018anti}. There is a taxonomy of defenses against Denial-of-Service (DoS) attacks that target ground networks \cite{mirkovic2004taxonomy}.
There is a survey of defenses against malware attacks that target terrestrial IT infrastructures  
\cite{ferdous2023review}. This is a systematization of defenses against cyber and electromagnetic attacks that target space infrastructures reported in academic literature \cite{remy2025sok}. There is a systematization of defenses against attacks that target space infrastructures via {\em signal}, {\em space platforms}, and {\em ground systems} \cite{pavur2022building}.
These taxonomies are different from ours because: (i) we consider defenses against counterspace attacks; {\color{black}(ii) we describe the entry and impact points of each attack}; (iii) we map defenses to the attacks that are thwarted by them; and (iv) we describe the {\em deployment point} for each defense.

\smallskip

\noindent{\bf Paper Outline}. 
Section \ref{sec:terminology} reviews preliminary knowledge.
Section \ref{sec:attack-taxonomy} describes our attack taxonomy.
Section \ref{sec:defense-taxonomy} presents our defense taxonomy. 
Section 
\ref{sec:limitations} discusses limitations of the present study. 
Section \ref{sec:conclusion} concludes the paper.

\section{Preliminaries}
\label{sec:terminology}

\begin{figure}[!htbp]
\centering
\includegraphics[width=.6\columnwidth]{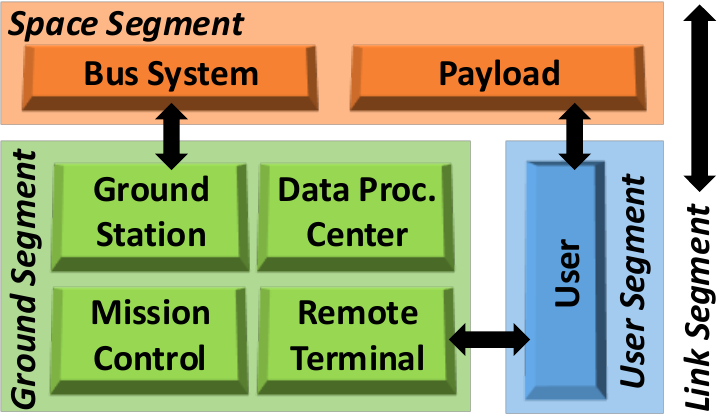}
\caption{System model of space infrastructures (adapted from \cite{remy2025sok}).}
\label{fig:simplified_system_model}
\end{figure}

\noindent{\bf Terminology and System Model}. We use the term {\em spacecraft} to represent any man-made object orbiting in space, 
such as satellites.
We adopt the  system model of {\em space infrastructures}  introduced in \cite{remy2025sok}, which consists of four levels of abstraction (high to low level): {\em mission}, {\em segment}, {\em component}, and {\em modules}. Examples of missions include bus management, Positioning, Navigation and Timing (PNT), scientific applications, 
and communications. 
Figure 
\ref{fig:simplified_system_model} highlights a simplified version of space infrastructures at the segment and component levels of abstraction. Specifically, the segment level of abstraction includes the {\em space}, {\em ground}, {\em user}, and {\em link} segments. The component level of abstraction includes the {\em bus system} and {\em payload} components in the space segment, the {\em user} component in the user segment, and the {\em data processing center}, {\em ground station}, {\em mission control}, and {\em remote terminal} components in the ground segment. We refer to \cite{remy2025sok} for more details.

In relation to the security of spacecraft, the term {\em compromised spacecraft} refers to a benign and victim spacecraft that is under the control of the attacker via cyber attacks.
The term {\em target spacecraft} means a spacecraft that  
an attacker is attempting to attack via counterspace, electromagnetic, or cyber means.
The term {\em malicious spacecraft} refers to a 
spacecraft that is commanded 
by an attacker.

To clearly describe the taxonomies, we use the term {\em space network} to describe a set of components in the space segment that can formulate communication networks (e.g., the payload components of satellites form a network, which is often manifested as a satellite network or constellation). 
We use the term {\em ground network} to describe a set of components in the ground segment that can form communication networks (e.g., the data processing centers can form a network, which does not include components of the user segment. 
Note that {\em ground networks} are typically connected to {\em space networks} through a ground station components module known as {\em gateways}.





\begin{figure}[!htbp]
\centering
\includegraphics[width=.9\columnwidth]{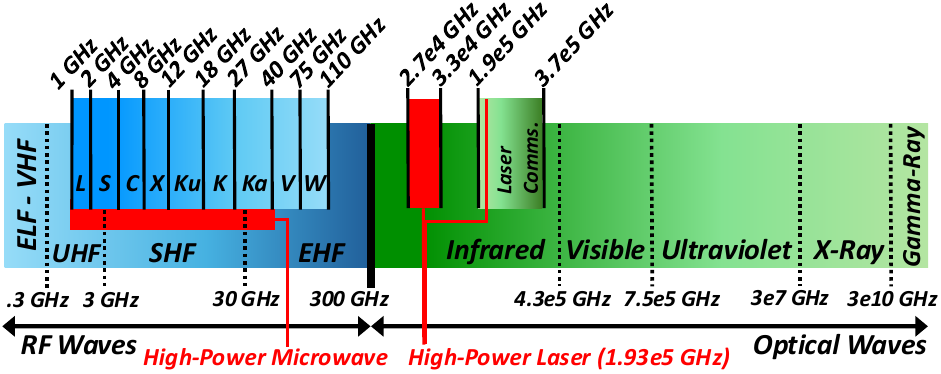}
\vspace{-.8em}
\caption{The electromagnetic wave spectrum 
(adapted from \cite{norgard2017electromagnetic, ITUfreqbands, IEEEfreqbands}). 
ELF (Extremely Low Frequency), VHF (Very High Frequency), UHF (Ultra High Frequency), SHF (Super High Frequency), and EHF (Extremely High Frequency) are
frequency bands in the Radiofrequency (RF) portion of the spectrum (i.e., RF waves). Infrared, visible, ultraviolet, X-ray, and gamma-ray are optical waves. 
Frequencies in red are often used by attackers to wage electromagnetic attacks, such as {\em disruption - high-power EM wave} attacks. 
}
\label{fig:EM_spectrum}
\end{figure}

In relation to spacecraft communications, we use the following standard terms.
The term {\em electromagnetic wave} denotes a radiation of electromagnetic energy, which is characterized by its frequency, wavelength, and
amplitude.
The term {\em electromagnetic spectrum} denotes a set of electromagnetic waves, often described by their wavelengths or frequencies.
A {\em frequency band} is a set of frequencies within an electromagnetic spectrum, which are used by transmitters and receivers. Figure \ref{fig:EM_spectrum} highlights a standardized nomenclature for 
frequency bands, such as the L, S, C, X, Ku, K, Ka, V, and W bands 
\cite{norgard2017electromagnetic, ITUfreqbands, IEEEfreqbands}.

%


A communication {\em channel} denotes a set of frequencies within a frequency band, and is described 
by its center frequency (also known as the {\em carrier frequency}) as well as its bandwidth (i.e., the number of frequencies the channel occupies in a frequency band). For instance, a .5 GHz bandwidth channel in the L band with a carrier frequency of 1.5 GHz means the channel expands from 1.25 GHz to 1.75 GHz. 
A channel can suffer from {\em noise}, and a large amount of noise can render the channel unusable.
The term {\em compromised channel} means a victim channel that is controlled by the attacker. The term {\em target channel} refers to a channel that the attacker is attempting to compromise but has yet to be compromised.

The term {\em signal} represents a use of an electromagnetic wave that carries data from a transmitter to a receiver through a channel, where both the transmitter and the receiver use an antenna to 
radiate 
and 
receive the signal, respectively \cite{wertz2011space}. 
A signal has a bandwidth.
The term {\em beam} describes the shape of the antenna radiation pattern. 
The term {\em signal footprint} describes the area in which the signal is received.
As highlighted in Figure \ref{fig:EM_spectrum}, a {\em Radiofrequency (RF) signal} is a signal that uses a RF wave, while 
an {\em optical signal} uses an optical wave.
A {\em downlink signal} is a signal where the transmitter is in a spacecraft and the receiver is in the ground segment. An {\em uplink signal} is a signal where the transmitter is in the ground segment and the receiver is in a spacecraft. An {\em Inter-Satellite Link (ISL) signal} is a signal where both the transmitter and receiver are in spacecraft. 

\noindent{\bf Threat Model (Attack Description)}. 
We categorize attacks against space infrastructures into: 
counterspace, electromagnetic, and cyber. Each attack is described by four attributes.
(i) The {\em attacker objective} (the ``what''), 
which can be mapped to one or multiple attack {\em tactics} in the ATT\&CK \cite{ATTCK} and/or SPARTA framework \cite{SPARTA}. (ii) The {\em attacker capability}, which describes the means that can be used by the attacker to achieve its objective and can be mapped to attack {\em techniques} (the ``how'') or attack {\em sub-techniques} (the detailed ``how'') as described in the ATT\&CK and SPARTA frameworks. (iii) The {\em entry point}, which describes 
the segment / component / module 
at which an attacker penetrates into a space infrastructure. (iv) The 
{\em impact point}, which describes the segment / component / module 
at which the effect of the attack is manifested
\cite{XuCNS2023,remy2025sok,remy2025quantifying}. Note that we propose differentiating entry points and impact points because 
the cyber attacks whose entry points and impact points are in the ground and/or user segment have already been accommodated by the cyber attacks against terrestrial networks; i.e., these attacks are not considered in the present paper because we focus on attacks whose entry and/or impact points are in space or link segment. 

\noindent{\bf Defense Description}. 
We categorize
defenses for space infrastructures into:
counterspace, electromagnetic, and cyber, which correspond to the three attack categories. Each defense is described by three attributes.
(i) The {\em defense type}, which is defined as preventive (i.e., aiming to prevent attacks from succeeding) and reactive {\color{black}(i.e., aiming to detect and respond to attacks)}.
(ii) The {\em defense mechanism} (i.e., defender ``how"), which describes how the defense 
works. 
(iii) The {\em deployment point}, which describes the segment / component / module 
where the defense mechanism is employed.


\section{The Attack Taxonomy}
\label{sec:attack-taxonomy}

We divide attacks against space infrastructures into three 
categories: 
counterspace, 
electromagnetic, 
and cyber.

\subsection{Counterspace Attacks}
\label{subsec:kinetic_tax}

\begin{figure*}[!htbp]
\centering
\includegraphics[width=.65\textwidth]{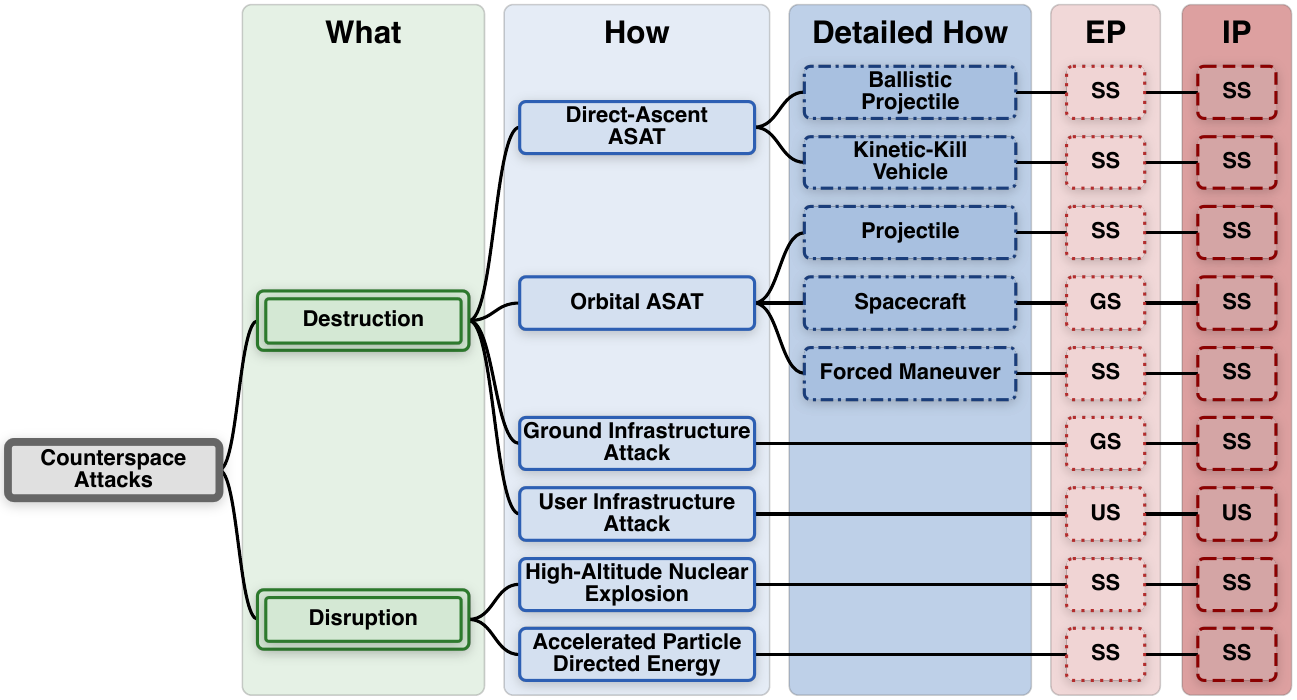}
\vspace{-.8em}
\caption{Taxonomy of counterspace attacks, describing their {\em attacker objectives} (the ``what'') in green boxes with double-lined borders, {\em attack capabilities} (the ``how'') in light-blue boxes with single-line borders, {\em attack sub-capabilities}
(the detailed ``how'') in dark-blue boxes with dot-dashed borders, {\em entry points} (EPs),
and {\em impact points} (IPs).
GS stands for Ground Segment, SS stands for Space Segment, and US stands for User Segment. 
}
\label{fig:counterspace_tree}
\end{figure*}

As highlighted in Figure \ref{fig:counterspace_tree}, this category can be further divided into two subcategories based on attacker objectives (i.e., the ``what''): 
{\em destruction} vs. {\em disruption} of spacecraft.


\subsubsection{Attacker Objective - Destruction Attacks}
These attacks use kinetic weapons/energy to produce a direct impact on spacecraft \cite{swope2024space,swope2025space}. They are divided into four 
attacker capabilities (the ``how''): Direct-Ascent Anti-Satellite (ASAT) attacks, Orbital ASAT attacks, Ground Infrastructure attacks, and User Infrastructure attacks.


\smallskip

\noindent{\bf (Counterspace Attacker Capability 1.1) Direct-Ascent ASAT Attacks}.
These attacks use malicious projectiles to attack spacecraft. The consequence is the destruction of victim spacecraft, which may incur many pieces of
debris
\cite{thiele2022investigating, kelso2007analysis}. Since projectiles need to complete an ascent to 
collide with the target spacecraft, attacking spacecraft at a higher altitude requires the attacker to spend a greater amount of energy. Only four nations have demonstrated the capability to target Low Earth Orbit (LEO) satellites: the United States  in 1985, reaching a 555 km altitude, and in 2008, reaching a 270 km altitude \cite{usASATs}; 
China in 2007, reaching an 865 km altitude \cite{weeden2020current}; India in 2019, reaching a 282 km altitude \cite{thiele2022investigating}; and Russia in 2021, reaching a 465 km altitude \cite{russianASAT}. 
These capabilities can be divided into two attacks (i.e., the detailed ``how''). 

{\bf (Counterspace Attacker Subcapability 1.1.1) Ballistic Projectile Attacks}.
In these attacks, the attacker uses a high-speed ballistic projectile to target a victim spacecraft, which is both the attack {\em entry point}
and the attack {\em impact point}  \cite{usASATs,weeden2020current, thiele2022investigating, russianASAT}. A ballistic 
projectile, such as a  
ballistic missile, often has a pre-calculated trajectory (i.e., the flight path) that resembles a parabola intercepting the target satellite trajectory \cite{harlin2007ballistic}. The flight path of a ballistic projectile is often described in three phases \cite{masters2013us}: (i) the {\em boost phase}, which is the initial thrust to achieve the proper ascension speed and trajectory; (ii) the {\em post-boost} phase, in which the projectile flies without thrust, namely freefall, relying on gravity and air friction; and (iii) the {\em impact phase}, in which the projectile collides with the victim spacecraft. To wage this attack, 
the attacker needs to have a
ground infrastructure for tracking target spacecraft
and launching projectiles.
This attack can be waged from the land, sea, or air (e.g., 1985 United States ASAT projectile was deployed from an F-15 fighter \cite{usASATs}). 

{\bf (Counterspace Attacker Subcapability 1.1.2) Kinetic-Kill Vehicle Attacks}.
In these attacks, the attacker uses a 
guided projectile (also known as Kinetic-Kill Vehicle, or KKV for short) to target a victim spacecraft, which is both the attack {\em entry point}
and the attack {\em impact point}  \cite{usASATs,weeden2020current, thiele2022investigating, russianASAT}. In contrast to ballistic projectiles, KKVs 
can maneuver during the impact phase \cite{wright2023hypersonic}. The requirement for waging this attack is the same as the requirement for launching the Ballistic Projectile Attack.

\smallskip

\noindent{\bf (Counterspace Attacker Capability 1.2) Orbital ASAT Attacks}.
These attacks target a 
spacecraft from a malicious spacecraft. The consequence of an orbital ASAT attack is the destruction of a victim spacecraft. These capabilities can be further divided into three attacks (i.e., the detailed ``how'').

{\bf (Counterspace Attacker Subcapability 1.2.1) Projectile Attack}.
This attack uses a malicious spacecraft to launch a projectile against a victim spacecraft, which is both the attack {\em entry point} and the attack {\em impact point}  \cite{russiaOrbitalASAT}. To our knowledge, only Russia has demonstrated projectile orbital ASAT capability in 2020 \cite{russiaOrbitalASAT}. 
To launch the attack, the attacker must have a ground infrastructure to 
track targeted spacecraft, be able to command the malicious spacecraft, 
and possess launch capabilities (i.e., launching malicious spacecraft into space and launching projectiles from their malicious spacecraft). 

{\bf (Counterspace Attacker Subcapability 1.2.2) Spacecraft Attack}. 
In this attack, 
the attacker is located in the ground segment and 
uses a compromised spacecraft to  
conjunct with a victim spacecraft. Note that only unintentional conjunctions between spacecraft have been publicly reported  (e.g., incurred by orbit miscalculation) \cite{NASAOrbitalCollision}.
To wage this attack, the attacker can leverage a signal-based attack or a cyber attack to compromise a spacecraft that has the capability to maneuver.

{\bf (Counterspace Attacker Subcapability 1.2.3) Forced Maneuver Attack}.
In this attack, the attacker is in the space and uses a malicious spacecraft to grab a target spacecraft, forcing the target spacecraft to maneuver to conjunct with another spacecraft, which is both the attack {\em entry point} and the attack {\em impact point}  \cite{chineseOrbitalASAT}. China has reportedly demonstrated this capability in 2022 by maneuvering a satellite to a graveyard orbit \cite{swope2025space}.
To wage this attack, the attacker must have a ground infrastructure for tracking target spacecraft, be able to command malicious spacecraft, and launch capabilities to send malicious spacecraft into space.

\smallskip

\noindent{\bf (Counterspace Attacker Capability 1.3) Ground Infrastructure Attacks}.
These attacks are waged by attackers on Earth
by using a projectile to attack a victim ground segment (e.g., an antenna in the ground station component, which is the attack {\em entry point}), while the space segment is typically 
the attack {\em impact point} \cite{swope2024space,swope2025space}. The consequence of these attacks is the destruction of a victim ground infrastructure, preventing the victim ground segment from communicating, 
commanding, and controlling the victim's spacecraft. To wage the attack, the attacker must have the capability to launch ballistic projectiles.

\smallskip

\noindent{\bf (Counterspace Attacker Capability 1.4) User Infrastructure Attacks}.
These attacks are waged by attackers on Earth by using a projectile to attack a victim user segment, such as a communications relay in the user component, which is both the attack {\em entry point} and the attack {\em impact point}. The consequence of these attacks is the destruction of a victim user segment, preventing the victim user segment from communicating with spacecraft.
To wage these attacks, the attacker must have the capability to launch projectiles. 


\subsubsection{Attacker Objective - Disruption Attacks}
These attacks use non-kinetic weapons/radiated energy 
\cite{swope2025space,falco2021security} 
to produce a direct physical impact on the target spacecraft. 
They can be further divided into two 
attacker capabilities (the ``how''): High-Altitude Nuclear Explosion and Accelerated Particle Directed Energy.

\smallskip

\noindent{\bf (Counterspace Attacker Capability 2.1) High-Altitude Nuclear Explosion Attacks}.
These attacks \cite{hess1964effects} are typically waged by an attacker on Earth which uses a nuclear weapon, that after detonation, emits a high amount of energy in multiple directions 
to attack targeted spacecraft, which are both the attack {\em entry point} and the attack {\em impact point}. These weapons are often detonated at high altitudes (i.e., above 160 km), 
emitting X-ray and gamma-ray energy as well as highly energized neutrons \cite{snyder2025effects}. Note that 
the atmospheric density decreases with altitude, where its minimum is often approximated at 160 km of altitude \cite{united1976us}. The consequence of these attacks is the destruction of the electronic components of victim spacecraft because 
X-ray and gamma-ray energy emissions supercharge the electronic equipment of victim spacecraft, and highly energized neutrons degrade the physical integrity of the electronic equipment of victim spacecraft \cite{snyder2025effects}. 
To wage these attacks, the attacker must have the capability
to launch ballistic projectiles.

\smallskip

\noindent{\bf (Counterspace Attacker Capability 2.2) Accelerated Particle Directed Energy Attacks}.
These attacks are typically waged by an attacker located in the space segment by radiating highly accelerated particles with no charge to attack a target spacecraft, while noting that charged particles are attracted by the inner and outer radiation belts---accumulations of extremely charged particles such as protons and electrons \cite{earthsRadiationBelts}. 
The victim spacecraft is both the attack {\em entry point} and the attack {\em impact point}. Note that 
accelerated particles lose energy when they interact with the particles suspended in the atmosphere 
\cite{zhou2014particle}. The consequence of these attacks is the destruction of a targeted spacecraft.
To wage these attacks, the attacker must have the capability to accelerate particles in space. 
It seems that the technology to accelerate particles as a weapon 
is still in early development, 
perhaps because it requires an extremely high amount of energy
\cite{zhou2014particle, liu2020space}.

\subsection{Electromagnetic Attacks}
\label{subsec:jamming_tax}

\begin{figure*}[!htbp]
\centering
\includegraphics[width=.65\textwidth]{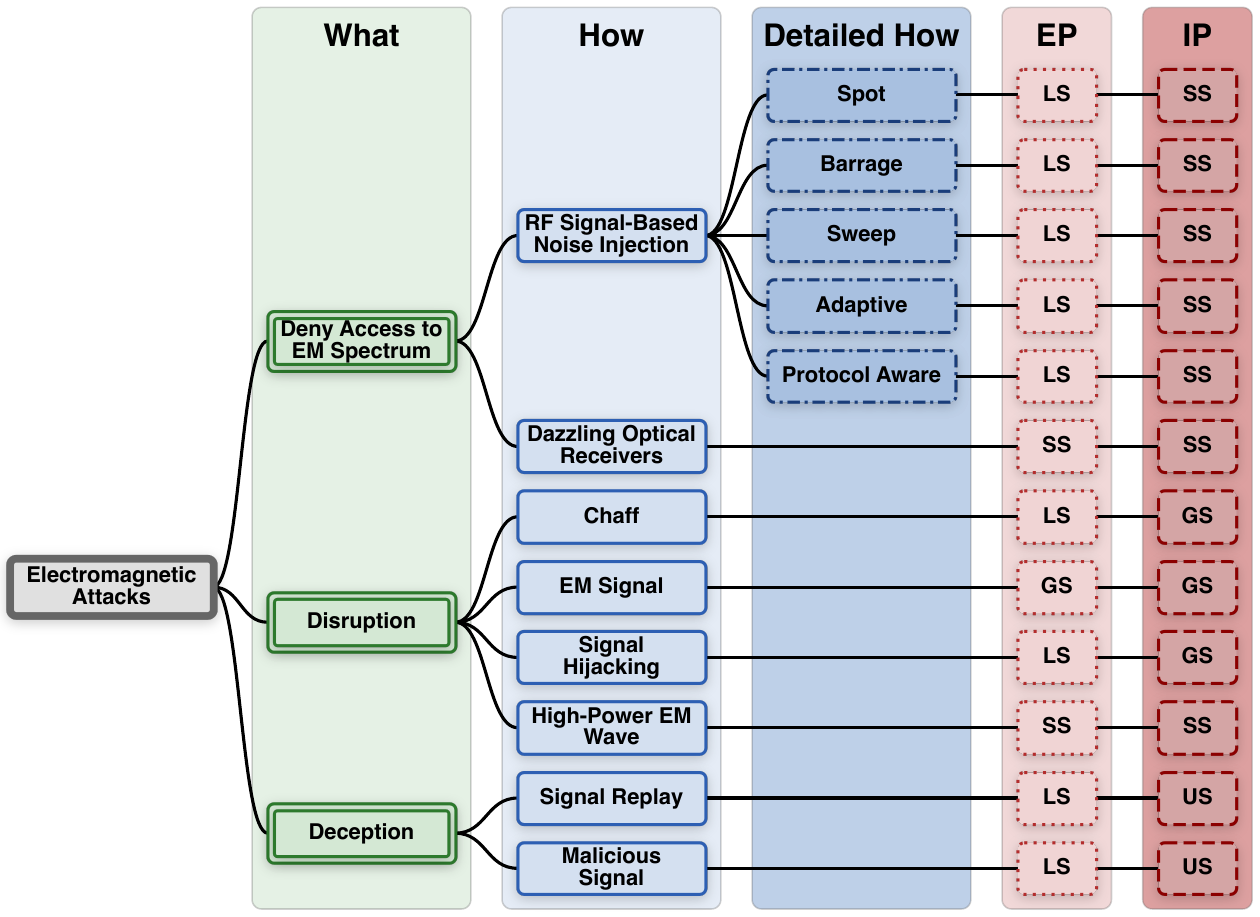}
\vspace{-.8em}
\caption{Taxonomy of electromagnetic attacks, describing their {\em attacker objective} (the ``what''),
{\em attacker capabilities} (the ``how''),
{\em attack sub-capabilities}
(the detailed ``how''),
attack {\em entry points} (EPs),
and attack {\em impact points} (IPs).
GS stands for Ground Segment, SS stands for Space Segment, US stands for User Segment, and LS stands for Link Segment. 
}
\label{fig:electromagnetic_tree}
\end{figure*}


As highlighted in Figure \ref{fig:electromagnetic_tree},  electromagnetic attacks can be further divided into three subcategories based on their attack objectives (the ``what''): {\em deny access to Electromagnetic (EM) spectrum} vs. {\em disruption} vs. {\em deception} of spacecraft.
%


\subsubsection{Attacker Objective - Deny Access to EM Spectrum Attacks}
These attacks use a malicious signal to prevent the targeted space segment 
from using the EM spectrum by clogging the frequency band or channel used between victim segments.
The consequence is that the victims cannot communicate.
These attacks can be divided into two attacker capabilities: 
RF Signal-Based Noise Injection, 
and Dazzling Optical Receivers attacks.


\smallskip

\noindent{\bf (Electromagnetic Attacker Capability 1.1) RF Signal-Based Noise Injection Attacks}. 
These attacks can be waged from the ground \cite{remy2025sok,jameel2018comprehensive,huo2017jamming,shahid2024taxonomy}, space \cite{falco2020satellites}, and air (e.g., drones), by injecting 
noise into the target channel or frequency band between a target receiver and a target transmitter. The attack {\em entry point} is the channel or frequency band, and the attack {\em impact point} is the victim receiver. 
The impact is that the target receiver cannot recover legitimate signals from the noise. 
These capabilities can be further divided into five attacks (the detailed ``how''). 

{\bf (Electromagnetic Attacker Subcapability 1.1.1) Spot Attack}. In this attack, the attacker constantly injects noise into the center frequency of the targeted channel. This attack is especially applicable to legacy spacecraft that operate only in some predetermined channel(s).

{\bf (Electromagnetic Attacker Subcapability 1.1.2) Barrage Attack}. This attack continuously injects noise into the center frequency of the targeted frequency band by using a signal that has the same bandwidth as the frequency band. This attack is especially applicable if the attacker wants to deny the entire frequency band that is used between a target receiver and a target transmitter. Note that even if the target transmitter and receiver change channels, the attack would still succeed because all channels in the frequency band are affected. 

{\bf (Electromagnetic Attacker Subcapability 1.1.3) Sweep Attack}. In this attack, the attacker injects noise into random frequencies in a target frequency band. The attacker randomly selects frequencies and duty cycles (i.e., the time intervals during which the attacker changes the frequency being targeted). This attack is especially applicable when the attacker only knows the frequency band used by the target sender and the target receiver.

{\bf (Electromagnetic Attacker Subcapability 1.1.4) Adaptive Attack}. This attack reacts to the victim transmission by adaptively choosing the frequency 
and duty cycles. 
This attack is applicable when the target transmitter and receiver have 
a synchronization mechanism, 
such as (i) shifting the frequency that is being used when noise is detected in the channel or (ii) stopping transmissions until 
the noise in the channel decreases.
This attack differs 
from the preceding three attacks because the attacker 
needs to eavesdrop on target transmissions.

{\bf (Electromagnetic Attacker Subcapability 1.1.5) Protocol Aware Attacks}. 
In these attacks,
the attacker selects duty cycles 
to inject noise into the key moments of the transmission \cite{salkield2025spacejam, hussain2014protocol}.
To wage this attack, the attacker must know the protocol used by the target transmitter and receiver.

\smallskip

\noindent{\bf (Electromagnetic Attacker Capability 1.2) Dazzling Optical Receiver Attacks}. 
These attacks can be waged by an attacker on Earth, using a
malicious optical signal (e.g., a laser signal using waves of wavelengths between 808nm and 847nm \cite{toyoshima2008ground})
to target a spacecraft with an optical receiver, which is both the attack {\em entry point} and the attack {\em impact point}.
Laser-based communications are commonly used between a ground station and a satellite in a Geosynchronous (GEO) orbit, where the relative position of the satellite to the ground station remains constant. It is worth mentioning that laser-based communications are degraded by 
atmospheric conditions such as clouds, rain, and by pointing errors \cite{toyoshima2008ground}, which are caused by the misalignment between the transmitter laser and the optical receiver. Moreover, the beam of a laser signal is substantially narrower than the beam of an RF signal, which makes tracking and pointing especially important. 
The consequence of these attacks is that the target spacecraft cannot receive uplink signals from the ground segment. 


\subsubsection{Attacker Objective - Disruption Attacks}
These attacks
use a reflecting surface to prevent legitimate signals from reaching the victim receiver in the ground segment.
These attacks can be further divided into three attacker capabilities: 
{\em Chaff}, {\em EM signal}, and {\em Signal Hijacking}.

\smallskip

\noindent{\bf (Electromagnetic Attacker Capability 2.1) Chaff Attacks}.
These attacks \cite{payne2006principles} can be waged by attackers on Earth, using a large number of small reflecting surfaces, or chaff, which resemble the shape of clouds, to target a certain type of ground segment receivers. The target signal is the attack {\em entry point} and the target radar receiver is the attack {\em impact point}.
Note that a radar is a ground segment infrastructure which is used to 
determine the distance to space objects (including spacecraft and debris). A radar transmits an EM signal to a space object and measures the time it takes for the scattered signal (i.e., the portion of the transmitted signal that bounces off the space object) to reach back to the radar. Note that the chaff cloud floating between the radar and the space object in question reflects the transmitted signal back to the radar. 
To wage these attacks, the attacker must be able to use a vehicle (e.g., aircraft \cite{kim2023chaff, pandey2013modeling}) to deploy a cloud of shafts.
The consequence of these attacks is that the target radar system cannot determine the orbit of the space object in question.

\smallskip

\noindent{\bf (Electromagnetic Attacker Capability 2.2) EM Signal Attacks}.
These attacks \cite{rawlins2022death} can be waged from the space segment and use a malicious signal to target 
a receiver in the ground segment, which is both the attack {\em entry point} and the attack {\em impact point}. These attacks are incorporated into our taxonomy because they can be waged from space. These attacks are applicable when the target receiver in the ground segment receives the legitimate signal from a GEO satellite.
To wage these attacks, the attacker must use a LEO satellite to transmit a malicious signal with a very low angular separation between the legitimate signal and the malicious signal. The consequence of these attacks is that the target receiver in the ground segment cannot differentiate the legitimate signal from the malicious signal. 

\smallskip

\noindent{\bf (Electromagnetic Attacker Capability 2.3) Signal Hijacking Attacks}. 
These attacks can be waged by an attacker on Earth using 
a malicious signal to target a receiver in the ground segment \cite{GlobalStarSimplexAttack,signalHijackingMaxHeadroom, manulis2021cyber, ear2023characterizing}, meaning the receiver is both the attack {\em entry point} and the attack {\em impact point}. The attacker transmits the malicious signal with the same characteristics as the legitimate signal, but with a higher power. These attacks are applicable to receivers that do not employ a signal authentication mechanism, such as satellite television stations,
because attacks succeed by using a higher power than the legitimate signal.


\smallskip

\noindent{\bf (Electromagnetic Attacker Capability 2.4) High-Power EM Wave Attacks}. 
These attacks \cite{pentagonlaser, liu2020space, falco2020satellites, swope2024space, swope2025space} can be waged by an attacker on Earth or in  space, sending a highly energized EM spectrum wave (e.g., laser or microwave) 
to a target 
spacecraft, which is both the attack {\em entry point} and the attack {\em impact point}. Note that the attacker can be on the ground because laser and microwave waves can traverse the atmosphere. To wage the attack, the attacker must have a ground segment infrastructure for tracking the victim spacecraft and an antenna for radiating high-energy EM waves. This attack can destroy the electronic modules of the victim spacecraft.


\subsubsection{Attacker Objective - Deception Attacks}
These attacks use a malicious signal
to deceive a target Global Navigation Satellite Systems (GNSS) receiver in the user segment. GNSS receivers, such as Global Positioning System (GPS) receivers, use multiple downlink navigation signals from satellites in GNSS constellations to calculate the location of a user on Earth by leveraging the time difference between the transmission and reception of navigation signals. When a GNSS receiver is booted, it acquires 
a minimum of four signals that carry enough power, but typically acquires a total of more 
than four. 
The time it takes the GNSS receiver to acquire four or more navigation signals is referred to as the Time To First Fix (TTFF). A GNSS receiver only uses the navigation signals acquired during the TTFF to calculate its position. If the power of an acquired navigation signal is below a threshold,
then the signal is not reliable and  
the receiver needs to calculate its location with the other signals while trying to re-acquire the unreliable signal. A receiver that cannot acquire, or re-acquire, at least four signals, cannot calculate its  
position (known as the ``no fix'' condition), forcing the receiver to reboot. When the receiver reboots, it discards previously acquired signals and acquires four or more new signals with high power. The attack makes a target receiver acquire its malicious navigation signal during the TTFF. If the target receiver has already acquired four or more signals, the attacker injects noise (via jamming) into the navigation signals used by the target receiver to impose the ``no fix'' condition, causing the receiver to reboot and get back to the TTFF \cite{tippenhauer2011requirements}. 
To wage these attacks, the attacker must be in proximity or in line of sight to the target receiver because the attacker needs to ensure that the power of the malicious navigation signal is higher than that of the legitimate navigation signal. 
The consequence is that a victim GNSS receiver obtains incorrect locations because of the received malicious navigation signals
\cite{altaweel2023gps}.
These attacks can be divided into two capabilities: 
{\em signal replay} and {\em malicious signal}. 

\smallskip

\noindent{\bf (Electromagnetic Attacker Capability 3.1) Signal Replay Attacks}.
These attacks
\cite{remy2025sok,wu2020spoofing} can be waged by an attacker on Earth, replaying a delayed legitimate navigation signal to the target receiver during the TTFF (i.e., when a receiver acquires
navigation signals). This means 
the signal (i.e., link segment) is 
the attack {\em entry point} and the user segment is the attack {\em impact point}. 

\smallskip

\noindent{\bf (Electromagnetic Attacker Capability 3.2) Malicious Signal Attacks}.
These attacks
\cite{remy2025sok,wu2020spoofing} 
can be waged by an attacker on Earth,
using a malicious navigation signal that is crafted as a legitimate signal against a target receiver. The signal (i.e., link segment) is the attack {\em entry point} and the user segment is the attack {\em impact point}. 

\subsection{Cyber Attacks}
\label{subsec:cyber_tax}

\begin{figure*}[!htbp]
\centering
\includegraphics[width=.65\textwidth]{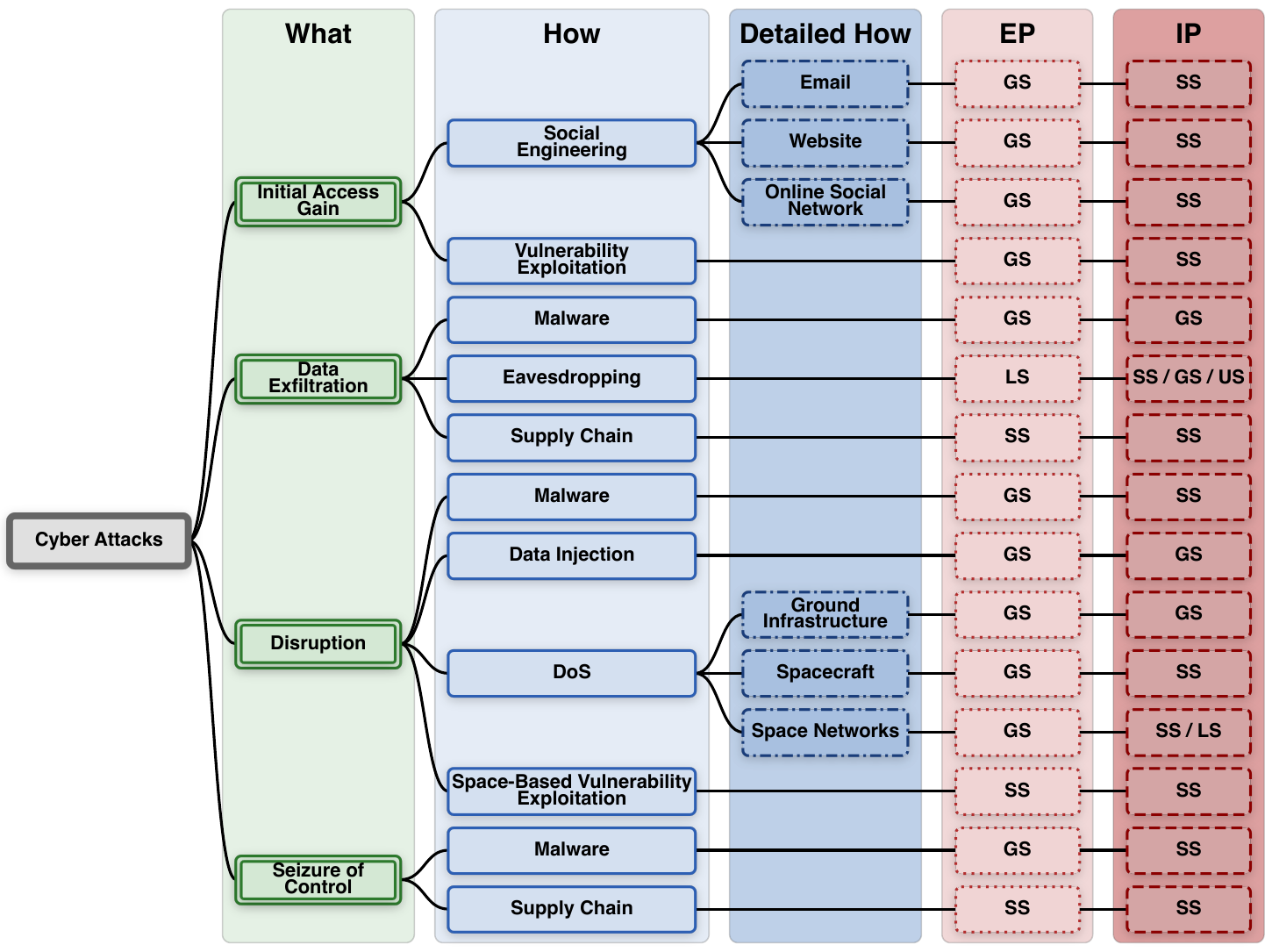}
\vspace{-.8em}
\caption{Taxonomy of cyber attacks against space infrastructures, describing their {\em attacker objective} (the ``what''),
{\em attack capabilities} (the ``how''),
{\em attack sub-capabilities} (the detailed ``how''),
attack {\em entry points} (EPs),
and attack {\em impact points} (IPs).
GS stands for Ground Segment, SS stands for Space Segment, and LS stands for Link Segment. 
}
\label{fig:cyber_tree}
\end{figure*}

As highlighted in Figure \ref{fig:cyber_tree}, cyber attacks against space infrastructures can be divided into four subcategories based on attacker objectives: {\em initial access gain} to ground segment, {\em data exfiltration} from ground segment, {\em disruption} of ground and space segments, and 
{\em seizure of control} of space segment.



\subsubsection{Attacker Objective - Initial Access Gain (to Ground Segment)}
\label{subsubsec:IAG}
These attacks can be waged by attackers using cyber capabilities to gain access to the target ground segment. They can further be divided into two 
capabilities: {\em social engineering} 
and {\em vulnerability exploitation}. 

\smallskip

\noindent{\bf ({\color{black} Cyber} Attacker Capability 1.1) Social Engineering Attacks}.
These attacks exploit psychological factors \cite{longtchi2024internet} of ground segment operators. 
The consequence is that the target/victim operator unknowingly provides the attacker with access to the ground segment components for which the operator is authorized. The attacker can use the operator's privileges to exfiltrate data, disrupt, or seize control of spacecraft.
This capability can be further divided into three attacks.

{\bf ({\color{black} Cyber} Attacker Subcapability 1.1.1) Email-based Social Engineering Attack}.
This attack 
\cite{longtchi2024internet, o2017insights}, 
sends malicious emails, such as spear phishing emails, to target operators, who may (e.g.) double-click 
the email attachments and thus get their computers compromised. 
There is evidence \cite{ear2023characterizing} showing that this attack has been waged by malicious actors to target aerospace companies 
to gain initial access to their ground segments and then laterally move to the space segment. 
In this attack, the attack {\em entry point} is the ground segment (more specifically, the operator or the component the operator is authorized to use) and the attack {\em impact point} is typically the space segment.


{\bf ({\color{black} Cyber} Attacker Subcapability 1.1.2) Website-based Social Engineering Attack} This attack 
\cite{longtchi2024internet} uses a malicious website (e.g., scareware, drive-by download, tabnabbing, or water holing website) to target ground segment operators. Even though there is no evidence that malicious actors have used this attack, the attack is possible.
In this attack, the attack {\em entry point} is the ground segment (more specifically, the operator or the component the operator is authorized to use), and the attack {\em impact point} is typically the space segment. 
To wage this attack, the attacker must be able to lure a victim to visit a malicious website.

{\bf ({\color{black} Cyber} Attacker Subcapability 1.1.3) Online Social Network-based Social Engineering Attack}.
This attack uses an online social network (e.g., app spoofing or smishing) to target a ground segment operator \cite{longtchi2024internet}. 
It is certainly possible even though we are not aware of any incident in the real world.
Similarly, the attack {\em entry point} is the ground segment, and the attack {\em impact point} is typically the space segment.

\smallskip

\noindent{\bf ({\color{black} Cyber} Attacker Capability 1.2) Vulnerability Exploitation Attacks}. 
These attacks are waged by attackers on Earth to exploit
vulnerabilities in a ground segment component, such as the remote terminal. A concrete example is the Terra satellite incident in 2008 \cite{ear2023characterizing}, whereby the attacker first compromises the ground segment and then laterally moves to the space segment. Thus, the attack {\em entry point} is the ground segment (more specifically, the vulnerable and compromised component) and the attack {\em impact point} is the space segment. 



\subsubsection{Attacker Objective - Data Exfiltration}
This subcategory of attacks uses cyber capabilities to exfiltrate data from the ground segment, the space segment, and the link segment (e.g., eavesdropping on the links). 
To exfiltrate data from the ground segment, the attacker must have access to it, 
which can be obtained by using an initial access gain attack mentioned above.
To exfiltrate data from the space segment, the attacker must penetrate into the spacecraft (e.g., via lateral movement).
To exfiltrate data from the link segment, such as a downlink signal, the attacker must be able to receive the signal, meaning the attacker is in the signal's footprint. These attacks can be further divided into three 
capabilities: malware, eavesdropping, and supply chain. 

\smallskip

\noindent{\bf ({\color{black} Cyber} Attacker Capability 2.1) Malware Attacks}.
These attacks \cite{ear2023characterizing, fritz2013satellite} 
can be waged by attackers 
using 
malicious software. 
For instance, 
in 2005, a malware infection 
at the Kennedy Space Center’s Vehicle Assembly Building spread to a NASA satellite control complex in Maryland and the Johnson Space Center in Houston, causing 20 gigabytes of data to be leaked to an attacker-controlled computer in Taiwan \cite{ear2023characterizing}. 
These attacks are accommodated in our taxonomy because the data leaked could be used to wage attacks against other segments (e.g., laterally moving to the space segment to exfiltrate data stored in spacecraft). The attack {\em entry point} is the ground segment, and the {\em impact point} is typically the ground segment, with further consequences to the space and link segments (e.g., an attacker gains information about the channel used between the ground station and the spacecraft and then injects noise into the channel). 

\smallskip

\noindent{\bf ({\color{black} Cyber} Attacker Capability 2.2) Eavesdropping Attacks}.
These attacks can 
collect data in transit between 
segments (e.g., between the space segment and the ground segment) \cite{ear2023characterizing,remy2025sok,pavur2020tale,willbold2023space}. These attacks listen to the target downlink 
signal, 
which is both the attack {\em entry point} and {\em impact point}. Note that the footprint of uplink signals is significantly smaller than that of downlink signals.
To wage these attacks, the attacker must have an RF receiver. The attack entry point is the link segment, but the impact point can be the space, ground, or user segment, depending on the nature of the exfiltrated data.




\smallskip

\noindent{\bf ({\color{black} Cyber} Attacker Capability 2.3) Supply Chain Attacks}.
These attacks can 
compromise components in the space segment (e.g.) the radio, 
before they are installed. The 
attack {\em entry point} and {\em impact point} are the space segment.
The consequence is that the attacker can obtain data from the compromised components and/or
legitimate downlink signals originating from the compromised spacecraft.


\subsubsection{Attacker Objective - Disruption}
This subcategory of attacks uses cyber capabilities to prevent the ground segment from 
controlling spacecraft 
or to prevent spacecraft from executing their missions. 
To wage these attacks, the attacker must have access to the ground segment, which can be achieved by using an initial access gain attack mentioned above. These attacks can be further divided into four 
attacker capabilities: {\em malware}, {\em data injection}, {\em DoS} (denial of service), and {\em vulnerability exploitation}. 


\smallskip

\noindent{\bf ({\color{black} Cyber} Attacker Capability 3.1) Malware Attacks}.
These attacks use malicious software (e.g., ransomware \cite{ear2023characterizing,falco2023wannafly, remy2025sok}) to compromised spacecraft. 
The attack {\em entry point} is typically in the victim ground segment, whereby the attacker uploads ransomware to run on spacecraft, which is the attack {\em impact point}. 
The consequence is a victim spacecraft can no longer be commanded or controlled without paying a ransom.


\smallskip

\noindent{\bf ({\color{black} Cyber} Attacker Capability 3.2) Data Injection Attacks}.
These attacks 
\cite{pavur2021detecting,pavur2019cyber,remy2025sok} 
inject malicious data into data processing center components in the ground segment, such as a Space Situational Awareness (SSA) repository, while noting that SSA data enables the ground segment to track locations of spacecraft. Both attack {\em entry point} and {\em impact point} are the ground segment. The consequence is that the victim ground segment cannot obtain the correct locations of spacecraft and other space objects (e.g., debris), which may lead to further consequences in the space segment (e.g., spacecraft conjunctions), which is why this attack is accommodated in our taxonomy.

\smallskip

\noindent{\bf ({\color{black} Cyber} Attacker Capability 3.3) DoS Attacks}.
These attacks use cyber capabilities to prevent a victim ground station from receiving or transmitting signals or to prevent spacecraft from operating properly. 
This subsubcategory can be divided into four attacks.

{\bf ({\color{black} Cyber} Attacker Subcapability 3.3.1) Ground Infrastructure Attack}.
This attack 
\cite{smailes2023dishing, remy2025sok} modifies the position of an antenna in the ground station component of a ground segment. Both attack {\em entry point} and {\em impact point} 
are 
the ground segment.
The consequence is that the victim ground segment cannot receive data from, or transmit data to, the space segment, which may suffer from further consequences, such as conjunction, which is why this attack is accommodated in our taxonomy.

{\bf ({\color{black} Cyber} Attacker Subcapability 3.3.2) Spacecraft Attack}.
This attack
attempts to prevents 
the operating system in the bus system component of a spacecraft from operating properly (e.g., executing a forkbomb to depleting the computer resources) \cite{thebarge2022developing, remy2025sok}. The attack {\em entry point} is in the ground segment, and the {\em impact point} is in the space segment. The consequence is that the operating system of a victim spacecraft is rendered unresponsive.

{\bf ({\color{black} Cyber} Attacker Subcapability 3.3.3) Space Networks Attack}.
This attack congests space networks with malicious 
traffic \cite{onen2004denial, usman2020mitigating}. 
Although it relies on Internet-based protocols (e.g., ICMP), {\color{black}the space network data is}
still modulated and encoded into RF signals, which are relayed between spacecraft in a space network. Moreover, limitations of space networks (e.g., longer transmission time) 
may amplify the effectiveness of the attack. DoS (Denial of Service) defense mechanisms, such as deploying additional resources \cite{mirkovic2004taxonomy}, are not applicable to space networks in a timely manner, as additional resources may translate to additional spacecraft.
The attack {\em entry point} is in the ground segment and {\em impact point} is in the space and link segments.
The consequence is that space network communication capabilities are degraded. 



\smallskip

\noindent{\bf ({\color{black} Cyber} Attacker Capability 3.4) Space-based Vulnerability Exploitation Attacks}.
These attacks 
\cite{hitefield2018exploiting, ear2023characterizing} 
exploit vulnerabilities in the bus system component of a target spacecraft, which controls the radio. 
For instance, the attacker sends a malformed packet to exploit a buffer overflow vulnerability in the software that controls the radio, preventing the radio from operating correctly. 
These attacks do not require the attacker to have initial access to the ground segment because 
the attacker can succeed by using a malicious signal. Both attack {\em entry point} and {\em impact point} 
are in the space segment. 
The consequence is that the victim spacecraft is unable to transmit or receive data.

\subsubsection{Attacker Objective - Seizure of Control}
These attacks use cyber capabilities to seize control of spacecraft. They may require the attacker to have access to the ground segment, which can be achieved by using an initial access gain attack described above. 
These attacks can be further divided into 
two 
attacker capabilities: {\em malware} and 
{\em supply chain}.


\smallskip

\noindent{\bf ({\color{black} Cyber} Attacker Capability 4.1) Malware Attacks}. 
These attacks \cite{remy2025sok} use malicious software to seize control of a target spacecraft. The attacker may upload and run malware in a spacecraft to modify the 
bus system component 
of the victim spacecraft (e.g., modifying the communication parameters to force the victim spacecraft to use a malicious channel). The attack {\em entry point} is typically the ground segment, and the attack {\em impact point} is the space segment.
The consequence is the attacker can 
remotely command and control the target spacecraft.



\smallskip

\noindent{\bf ({\color{black} Cyber} Attacker Capability 4.2) Supply Chain Attacks}.
These attacks can compromise (e.g.) a bus system component before it is installed in a spacecraft, such as the radio used to receive commands from the ground station in the ground segment. Both attack {\em entry point} and {\em impact point} are the space segment. 
Note that this attack does not require the attacker to have initial access to the ground segment because 
the attacker can succeed by using a malicious signal.
The consequence is that the attacker can remotely command and control a victim spacecraft via the compromised communications modules.

\section{The Defense Taxonomy}
\label{sec:defense-taxonomy}

Corresponding to the three categories of attacks, there are three categories of defenses: defenses against counterspace attacks or {\em counterspace defenses}, {\em electromagnetic defenses}, and {\em cyber defenses}.

\subsection{Counterspace Defenses}

\begin{figure*}[!htbp]
\centering
\includegraphics[width=.65\textwidth]{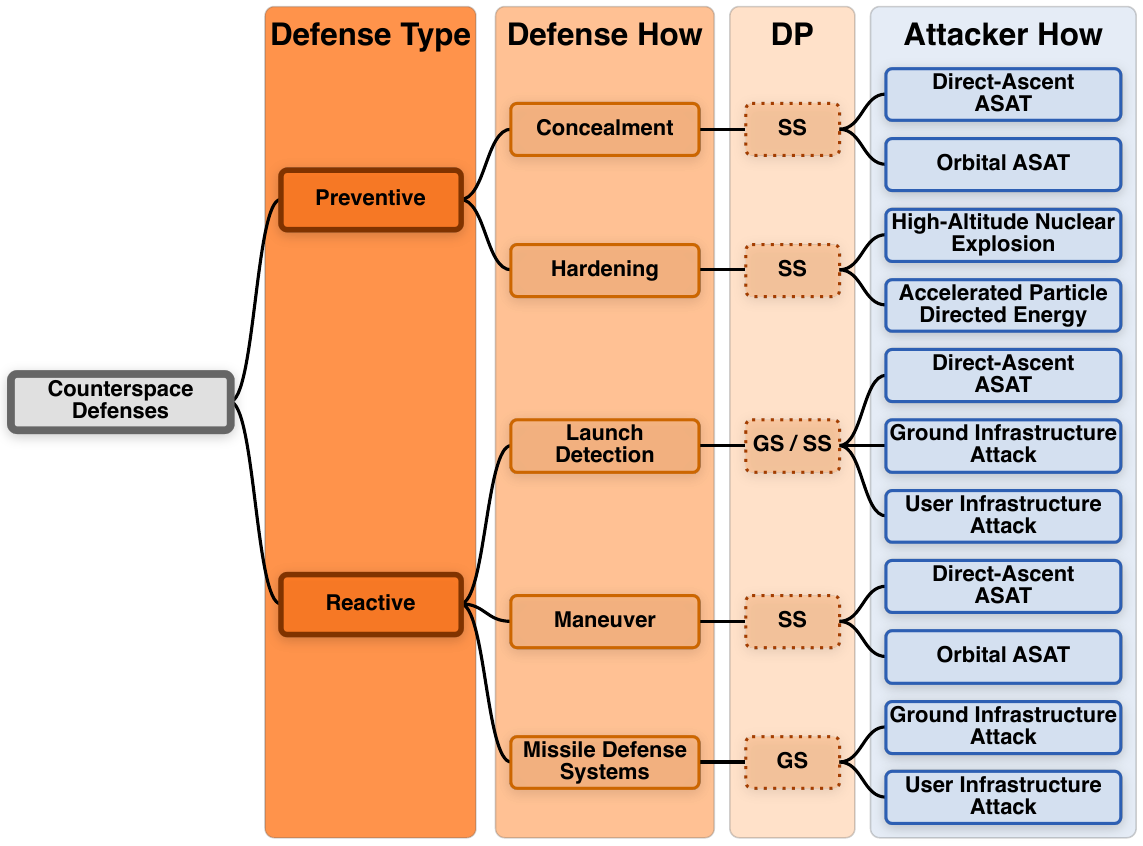}
\vspace{-.8em}
\caption{Taxonomy of counterspace defenses for space infrastructures, describing the {\em defense type}, {\em defense mechanisms} (i.e., defender ``how''), defense {\em deployment points} (DPs), and the 
{\em attack capabilities} (i.e., attacker ``how'') that can be thwarted by a defense. 
GS stands for Ground Segment, and SS stands for Space Segment. 
}
\label{fig:counterspace_def_tree}
\end{figure*}

As highlighted in Figure \ref{fig:counterspace_def_tree}, counterspace defenses can be divided into two types: 
{\em preventive} and {\em reactive} defenses. 

\subsubsection{Preventive Defenses}
These defenses aim to prevent a victim spacecraft from being detected and impacted by energy radiation. These defenses include
{\em concealment} and {\em hardening}.

\smallskip

\noindent{\bf ({\color{black} Counterspace Preventive Defenses}
1) Concealment}.
These defenses aim to prevent {\em direct-ascent ASAT} and {\em orbital ASAT} attacks by disrupting adversarial tracking of target spacecraft \cite{zhou2021integrated,zhao2025ultra,sun2022stealthy,reiter2020spacecraft,pavur2021detecting}. To disrupt tracking, a victim spacecraft can: (i) use anti-radar coatings, namely special materials, 
to make a spacecraft reflect or absorb energy irregularly, while leveraging satellite shapes to prevent radar systems on the gorund from tracking the victim spacecraft \cite{zhou2021integrated,zhao2025ultra,sun2022stealthy}, in a fashion similar to stealthy aircraft; (ii) perturb the orbit of a satellite \cite{reiter2020spacecraft}; and (iii) disguise a victim spacecraft as debris in public space objects location databases \cite{pavur2021detecting}. 
The defense {\em deployment point} is in the space segment. This defense draws results from the aerospace community \cite{bera2025overview} and is in early development. 


\smallskip

\noindent{\bf ({\color{black} Counterspace Preventive Defenses}
2) Hardening}.
These defenses aim to prevent 
the damage caused by 
{\em high-altitude nuclear explosion} and {\em accelerated particle directed energy} attacks by hardening spacecraft to deflect or absorb the 
energy radiation emitted from counterspace attacks \cite{wang2025review}. The defense {\em deployment point} is in the space segment. 







\subsubsection{Reactive Defenses}
These defenses aim to detect counterspace attacks. These defenses include  
{\em launch detection},
{\em maneuver}, and {\em missile defense systems}.


\smallskip

\noindent{\bf ({\color{black} Counterspace Reactive Defenses}
1) Launch Detection}.
These defenses, also known as {\em missile warning systems}, aim to detect when {\em direct-ascent ASAT}, {\em ground infrastructure}, and {\em user infrastructure} 
attacks are launched, which permits detection and tracking of malicious projectiles. 
The {\em deployment point} is in the ground segment (i.e., radar systems and optical cameras  
on Earth) or in the space segment, such as remote sensing payloads (e.g., infrared sensors), which can be used to measure the amount of heat released from a launch vehicle.

\smallskip

\noindent{\bf ({\color{black} Counterspace Reactive Defenses}
2) Maneuver}.
These defenses aim to maneuver 
target spacecraft away from the impact points of counterspace attacks in the space segment, such as {\em high-altitude nuclear explosion} and {\em accelerated particle directed energy} \cite{fetter1988protecting, yates2008systematic,shabbir2018counterspace}. 
The defense {\em deployment point} is in the space segment. These defenses are reactive because they require 
the defender to possess early-warning capabilities in detecting counterspace attacks and would demand substantial fuel consumption. 

\smallskip

\noindent{\bf ({\color{black} Counterspace Reactive Defenses}
3) Missile Defense Systems}.
These defenses react to {\em ground infrastructure} and {\em user infrastructure} attacks by using kinetic weapons to intercept malicious projectiles and thus preventing the malicious projectile from impacting \cite{shapir2013lessons, goldenDome}. 
The defense {\em deployment point} is in the ground segment. These defenses are reactive because they requires detection of malicious projectiles.

\subsection{Electromagnetic Defenses}
\begin{figure*}[!htbp]
\centering
\includegraphics[width=.65\textwidth]{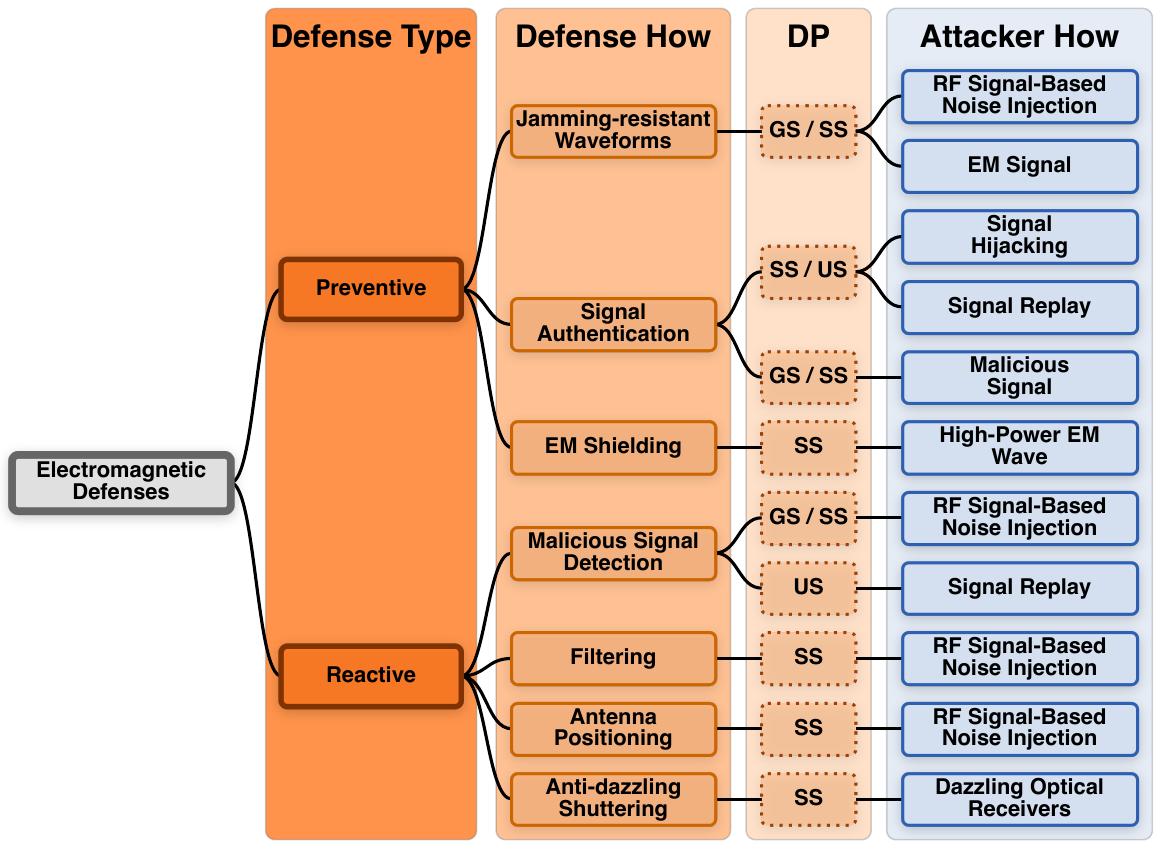}
\vspace{-.8em}
\caption{Taxonomy of counterspace defenses for space infrastructures, describing the {\em defense type}, {\em defense mechanisms} (i.e., defender ``how''), defense {\em deployment points} (DPs), and the 
{\em attack capabilities} (i.e., attacker ``how'') that can be thwarted by a defense. 
GS stands for Ground Segment, and SS stands for Space Segment. 
}
\label{fig:electromagnetic_def_tree}
\end{figure*}

As highlighted in Figure \ref{fig:electromagnetic_def_tree}, electromagnetic defenses has two types: 
{\em preventive} and {\em reactive}.

\subsubsection{Preventive Defenses}
These defenses aim to prevent 
jamming attacks, spoofing attacks, 
and physical damage to satellites. 
These defenses include {\em jamming-resistant waveforms},
{\em signal authentication}, and {\em EM shielding}.

\smallskip

\noindent{\bf ({\color{black}Preventive Electromagnetic Defenses}
1) Jamming-resistant Waveforms}.
These defenses aim to prevent {\em RF signal-based noise injection} and {\em EM signal} attacks by modifying the parameters of the communication channel used by the victim spacecraft \cite{jia2018anti, querol2017real}. The defense {\em deployment point} is in the ground and space segments, where the signal is shaped before transmission.


\smallskip

\noindent{\bf ({\color{black}Preventive Electromagnetic Defenses}
2) Signal Authentication}.
These defenses 
aim to prevent {\em signal hijacking} and {\em malicious signal} attacks by authenticating signals based on (i) physical layer security mechanisms that analyze physical parameters of the signal (e.g., the angle of arrival) \cite{vazquez2019rf,schraml2021multiuser,kalantari2015multibeam,cao2021noma,hayashi2020poisson}, or (ii) cryptography-based authentication (e.g., navigation message authentication) 
\cite{kerns2014blueprint,anderson2017chips,fernandez2014design, fernandez2023semi, curran2016message,wu2018ecdsa,pavur2021qpep,huwylerqpep,yang2019anafra,oligeri2023pastai}. The defense {\em deployment point} is in the ground station component of the ground segment, and in the bus system component of the space segment.

\noindent{\bf ({\color{black}Preventive Electromagnetic Defenses}
3) EM Shielding}. 
These defenses aim to reduce the impact of {\em high-power EM wave} attacks by shielding the victim spacecraft \cite{williams2020deployed,von2009composite}. 
The {\em deployment point} is in the space segment. 
Note that these defenses can be coupled with other defenses such as {\em maneuver} and {\em antenna positioning}.

\subsubsection{Reactive Defenses}
These defenses aim to detect malicious signals, such as the ones used in jamming attacks (i.e., {\em RF signal-based noise injection}) and spoofing (i.e., {\em signal replay}, and {\em malicious signal}), filter input signals, change antenna position of the victim spacecraft, and reduce impact of attacks on the payload of a victim spacecraft. 
These defenses include 
{\em malicious signal detection}, {\em filtering}, {\em antenna positioning}, and {\em anti-dazzling shuttering}.

\smallskip

\noindent{\bf ({\color{black}Reactive Electromagnetic Defenses}
1) Malicious Signal Detection}. 
These defenses aim to  detect malicious signals used in {\em RF signal-based noise injection} attacks \cite{borio2016jammer, querol2017real} (e.g., jamming against a victim spacecraft) and {\em signal replay} attacks 
\cite{wesson2017gnss,schmidt2020gps,humphreys2013detection,liu2023probabilistic,jovanovic2014multi,falletti2021performance,liu2021stars,lo2018robust,ceccato2021generalized,ceccato2018exploiting,jansen2016multi,psiaki2013gps,jansen2018crowd,clements2023dual,lachapelle2021orbital,spanghero2023detecting,spanghero2022high,spanghero2020authenticated,zhang2020protecting} 
(e.g., GNSS spoofing against victim users). 

To detect malicious signals used in {\em RF signal-based noise injection} attacks, the victim spacecraft can deploy a defense mechanism to analyze the signal-to-noise ratio of the malicious signal at different points of the victim orbit, which permits localization of the jammer's location \cite{borio2016jammer}. The defense {\em deployment point} is in the space segment.


To detect malicious signals used in {\em signal replay} attacks, the victim user can deploy a defense mechanism to analyze:
(i) the malicious signal power 
\cite{wesson2017gnss,schmidt2020gps,humphreys2013detection,liu2023probabilistic}; (ii) the malicious signal distance \cite{jovanovic2014multi,falletti2021performance}; (iii) the malicious signal direction of arrival \cite{liu2021stars,lo2018robust}; (iv) the victim user's
GNSS receiver data, 
e.g., comparing the position measurements produced by internal sensors 
and the position obtained from the malicious GNSS signal 
\cite{ceccato2021generalized, ceccato2018exploiting}; 
(v) location data derived from external sources, such as WIFI and cellular signals, which can be compared with the position obtained from the malicious GNSS signal 
\cite{jansen2016multi,psiaki2013gps,liu2023probabilistic,jansen2018crowd,clements2023dual,lachapelle2021orbital}; or (vii) time derived from external sources, such as the Network Time Protocol, which can be compared with the time obtained from the malicious GNSS signal 
\cite{spanghero2023detecting,spanghero2022high,spanghero2020authenticated,zhang2020protecting}. The {\em deployment point} of these defenses is in the user segment.

\smallskip

\noindent{\bf ({\color{black}Reactive Electromagnetic Defense}
2) Filtering}.
This defense aims to reduce the impact of {\em RF signal-based noise injection} attacks by analyzing the malicious jamming signal and filtering out the injected noise \cite{boriotracking}. The {\em deployment point} is in the space segment.

\smallskip

\noindent{\bf ({\color{black}Reactive Electromagnetic Defense}
3) Antenna Positioning}.
This defense aims to reduce the impact of {\em RF signal-based noise injection} attacks by modifying the position of the victim spacecraft antenna \cite{sun2022anti}. They assume there is an angular separation between the legitimate signal and the malicious signal, meaning that the victim antenna is positioned in the direction of the legitimate signal. The {\em deployment point} is in the space segment.

\smallskip

\noindent{\bf ({\color{black}Reactive Electromagnetic Defenses}
4) Anti-dazzling Shuttering}.
These defenses aim to reduce the impact of {\em dazzling optical receivers} attacks by closing the victim spacecraft's remote sensing payload shutter \cite{pollock1993infrared} or by modifying the position of the victim satellite \cite{trimoreau2022improving} to prevent the optical signal from damaging the sensing payload. The defense {\em deployment point} is in the space segment.

\subsection{Cyber Defenses}
\begin{figure*}[!htbp]
\centering
\includegraphics[width=.65\textwidth]{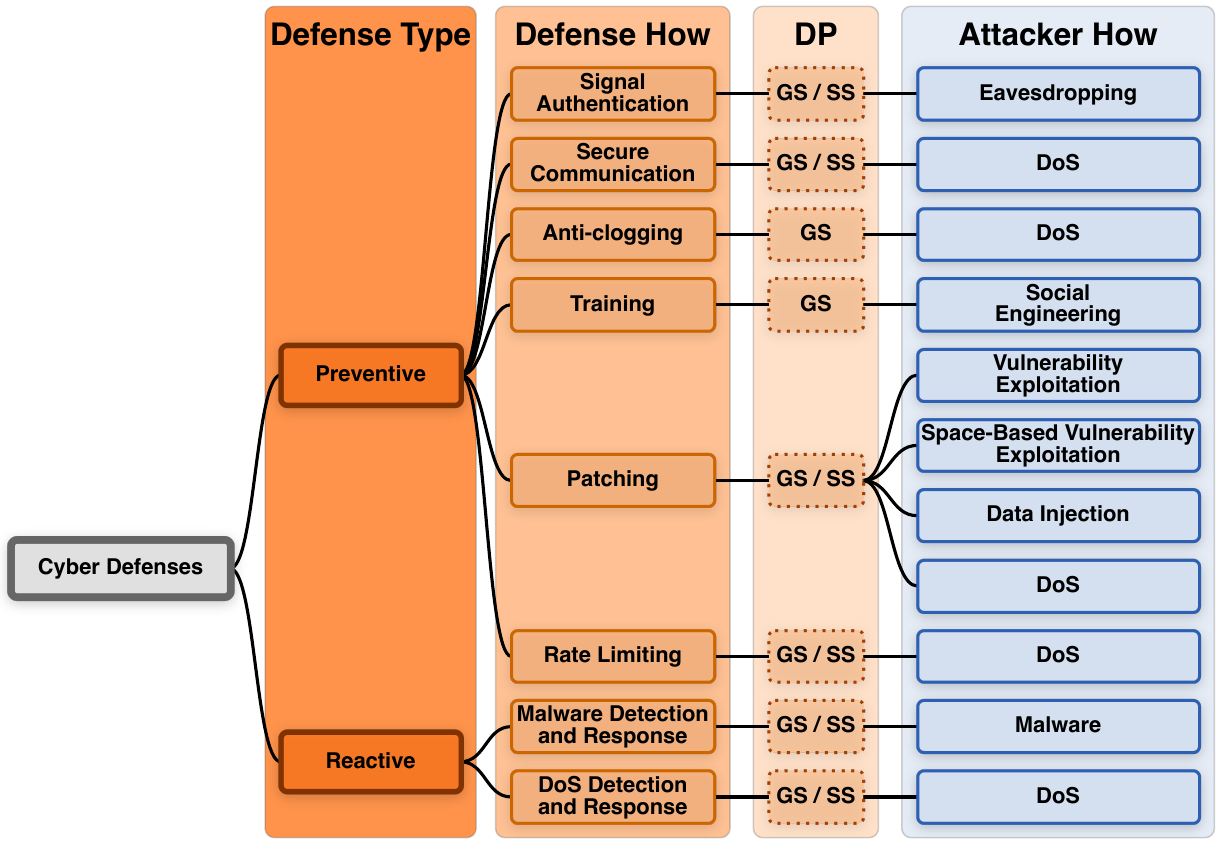}
\vspace{-.8em}
\caption{Taxonomy of counterspace defenses for space infrastructures, describing the {\em defense type}, {\em defense mechanisms} (i.e., defender ``how''), defense {\em deployment points} (DPs), and the 
{\em attack capabilities} (i.e., attacker ``how'') that can be thwarted by a defense. 
GS stands for Ground Segment, and SS stands for Space Segment. 
}
\label{fig:cyber_def_tree}
\end{figure*}

As highlighted in Figure \ref{fig:cyber_def_tree}, cyber defenses can be divided into two types: 
{\em preventive}, and {\em reactive} 
defenses. 

\subsubsection{Preventive Defenses}
These defenses aim to prevent disruption attacks, such as {\em DoS}, from impacting space networks, and eavesdropping attacks. 
These defenses include 
{\em secure communication}, {\em authentication}, {\em anti-clogging}, {\em training}, {\em patching}, and {\em rate limiting}.

\smallskip

\noindent{\bf ({\color{black}Preventive Cyber Defenses}
1) Secure Communication}.
These defenses aim to prevent {\em DoS - space networks} attacks by using communication protocols that are designed to prevent depletion of network resources \cite{mirkovic2004taxonomy}. Note that ground networks, which rely on Internet communication protocols, are not designed to resist DoS attacks
(e.g., TCP SYN flood attack \cite{salunkhe2017analysis,dang2019sdn}). The {\em deployment point} is in the ground and space segments. 

\smallskip

\noindent{\bf ({\color{black}Preventive Cyber Defenses}
2) Signal Authentication}.
These defenses
\cite{vazquez2019rf,schraml2021multiuser,kalantari2015multibeam,cao2021noma,hayashi2020poisson,kerns2014blueprint,anderson2017chips,fernandez2014design, fernandez2023semi, curran2016message,wu2018ecdsa,pavur2021qpep, huwylerqpep,yang2019anafra,oligeri2023pastai} prevent {\em eavesdropping} 
attacks by authenticating signals based on: (i) physical layer security \cite{vazquez2019rf,schraml2021multiuser,kalantari2015multibeam,cao2021noma,hayashi2020poisson}, 
and (ii) cryptography-based authentication 
\cite{kerns2014blueprint,anderson2017chips,fernandez2014design, fernandez2023semi, curran2016message,wu2018ecdsa,pavur2021qpep, huwylerqpep,yang2019anafra,oligeri2023pastai}. 
The {\em deployment point} is in the ground segment ground station component, and in the space segment bus system component.

\smallskip

\noindent{\bf ({\color{black}Preventive Cyber Defense}
3) Anti-clogging}.
This defense 
aims to prevent {\em DoS - space networks} attacks by authenticating the sender of the space network traffic to the receiver prior to the transmission of data \cite{onen2004denial}.
The {\em deployment point} is in the ground segment. 

\noindent{\bf ({\color{black}Preventive Cyber Defense}
4) Training}. 
This defense aims to reduce the impact of {\em social engineering attacks} by raising the awareness of employees against these attacks \cite{zaoui2024comprehensive}. The {\em deployment point} is in the ground segment. 

\smallskip

\noindent{\bf ({\color{black}Preventive Cyber Defense}
5) Patching}. 
 This defense aims to reduce the impact of {\em vulnerability exploitation}, {\em disruption - DoS - ground infrastructure}, and {\em disruption - DoS - spacecraft} attacks by eliminating vulnerabilities in victim components. The {\em deployment point} is in the ground and space segments.

\smallskip

\noindent{\bf ({\color{black}Preventive Cyber Defense}
6) Rate Limiting}. 
This defense aims to reduce the impact of {\em DoS - space networks} attacks by limiting the victim space network traffic \cite{mirkovic2004taxonomy}. 
The {\em deployment point} is in the remote terminal component of the ground segment and/or in the payload component of the space segment.

\subsubsection{Reactive Defenses}
These defenses aim to detect, and respond to, cyber attacks, such as malware and DoS attacks.  
These defenses include 
{\em malware detection and response}, 
and {\em DoS detection and response}.

\smallskip

\noindent{\bf ({\color{black}Reactive Cyber Defenses}
1) Malware Detection and Response}. These defenses aim to detect \cite{ferdous2023review}, and respond to \cite{oz2022survey}, {\em malware} attacks. 
Detection can be based on malware signatures
or malware behaviors \cite{ferdous2023review}.
The {\em deployment point} is in the ground segment (i.e., to detect {\em data exfiltration - malware} attacks) and/or in the space segment (e.g., 
bus system or payload components) to detect {\em disruption and seizure of control - malware} attacks. Responses to malware detected in the space segment, such as anti-virus software 
and Endpoint Detection and Response (EDR) defense mechanisms, aim to clean up the compromises but have not been duly investigated.

\smallskip


\smallskip

\noindent{\bf ({\color{black}Reactive Cyber Defenses}
2) DoS Detection {\color{black}and Response}}.
These defenses aim to detect, {\color{black}and respond to}, {\em DoS - space networks} attacks \cite{mirkovic2004taxonomy}. Detection can be based on known attack patterns
or anomaly detection \cite{mirkovic2004taxonomy}.
The {\em deployment point} is in the ground station component of the ground segment and/or in the bus system component of the space segment. Responses to DoS attacks include re-route space network traffic and reactive use of {\em rate limiting} mechanisms 
(i.e., rate limiting can be used for preventive and reactive defense purposes).

\smallskip

\section{Limitations}
\label{sec:limitations}

This study has two limitations, which should be addressed in future studies. First, we focus on technical attacks without considering organizational, regulatory, and legal risks. This is reasonable because we align our taxonomy to the ATT\&CK and SPARTA frameworks and the threat models described in the academic literature, which deal with technical capabilities. 
Second, some attack {\em entry points} and {\em impact points}, as well as some defense {\em deployment points} are described at a coarse granularity, meaning at the segment or component level of abstraction rather than the fine-grained module level in the terminology of the space infrastructure models. This is reasonable because attacks are often described at the {techniques} or {sub-techniques} level of abstraction rather than the {procedure} level of abstraction in the terminology of ATT\&CK and SPARTA frameworks. Nevertheless, the taxonomy can be easily extended to describe specific attack {\em procedures} when such information is available.

\section{Conclusion}
\label{sec:conclusion}
We have presented a taxonomy of space infrastructures attacks and defenses. The attack taxonomy includes counterspace attacks, electromagnetic attacks, and cyber attacks, where every attack is characterized by its entry point and impact point. 
The defense taxonomy correspondingly includes counterspace, electromagnetic, and cyber defenses, where each defense is characterized by its deployment point. 
The limitations discussed above represent interesting future research directions. We hope this paper will inspire a community effort to refine 
the taxonomy towards a widely-adoptable taxonomy.
 



\bibliographystyle{IEEEtran}

\begin{thebibliography}{100}
\providecommand{\url}[1]{#1}
\csname url@samestyle\endcsname
\providecommand{\newblock}{\relax}
\providecommand{\bibinfo}[2]{#2}
\providecommand{\BIBentrySTDinterwordspacing}{\spaceskip=0pt\relax}
\providecommand{\BIBentryALTinterwordstretchfactor}{4}
\providecommand{\BIBentryALTinterwordspacing}{\spaceskip=\fontdimen2\font plus
\BIBentryALTinterwordstretchfactor\fontdimen3\font minus \fontdimen4\font\relax}
\providecommand{\BIBforeignlanguage}[2]{{%
\expandafter\ifx\csname l@#1\endcsname\relax
\typeout{** WARNING: IEEEtran.bst: No hyphenation pattern has been}%
\typeout{** loaded for the language `#1'. Using the pattern for}%
\typeout{** the default language instead.}%
\else
\language=\csname l@#1\endcsname
\fi
#2}}
\providecommand{\BIBdecl}{\relax}
\BIBdecl

\bibitem{spacefoundation2022}
{Space Foundation Editorial Team}, ``Space foundation releases the space report 2022 q2 showing growth of global space economy,'' \url{https://www.spacefoundation.org/2022/07/27/the-space-report-2022-q2/}.

\bibitem{spacefoundation2023}
{Space Foundation Editorial Team}, ``Space foundation releases the space report 2023 q2, showing growth of global space economy to \$546b,'' \url{https://www.spacefoundation.org/2023/07/25/the-space-report-2023-q2/}.

\bibitem{remy2025sok}
J.~L.~C. Remy, E.~Ear, C.~Chang, A.~Feffer, and S.~Xu, ``Sok: Space infrastructures vulnerabilities, attacks and defenses,'' in \emph{2025 IEEE Symposium on Security and Privacy (SP)}.\hskip 1em plus 0.5em minus 0.4em\relax IEEE, 2025, pp. 1028--1046.

\bibitem{falco2021security}
G.~Falco and N.~Boschetti, ``A security risk taxonomy for commercial space missions,'' in \emph{ASCEND 2021}, 2021, p. 4241.

\bibitem{ceccato2021generalized}
M.~Ceccato, F.~Formaggio, N.~Laurenti, and S.~Tomasin, ``Generalized likelihood ratio test for gnss spoofing detection in devices with imu,'' \emph{IEEE TIFS}, 2021.

\bibitem{schmidt2016survey}
D.~Schmidt, K.~Radke, S.~Camtepe, E.~Foo, and M.~Ren, ``A survey and analysis of the gnss spoofing threat and countermeasures,'' \emph{ACM CSUR}, 2016.

\bibitem{xiao2018secure}
Y.~Xiao, J.~Liu, Y.~Shen, X.~Jiang, and N.~Shiratori, ``Secure communication in non-geostationary orbit satellite systems: A physical layer security perspective,'' \emph{IEEE Access}, 2018.

\bibitem{li2019physical}
B.~Li, Z.~Fei, C.~Zhou, and Y.~Zhang, ``Physical-layer security in space information networks: A survey,'' \emph{IEEE IoT-J}, 2019.

\bibitem{guo2021survey}
H.~Guo, J.~Li, J.~Liu, N.~Tian, and N.~Kato, ``A survey on space-air-ground-sea integrated network security in 6g,'' \emph{IEEE Communications Surveys \& Tutorials}, 2021.

\bibitem{tedeschi2022satellite}
P.~Tedeschi, S.~Sciancalepore, and R.~Di~Pietro, ``Satellite-based communications security: A survey of threats, solutions, and research challenges,'' \emph{Computer Networks}, 2022.

\bibitem{meng2022survey}
L.~Meng, L.~Yang, W.~Yang, and L.~Zhang, ``A survey of gnss spoofing and anti-spoofing technology,'' \emph{Remote Sensing}, 2022.

\bibitem{pavur2022building}
J.~Pavur and I.~Martinovic, ``Building a launchpad for satellite cyber-security research: lessons from 60 years of spaceflight,'' \emph{Journal of Cybersecurity}, 2022.

\bibitem{yuan2023authenticating}
M.~Yuan, X.~Tang, and G.~Ou, ``Authenticating gnss civilian signals: A survey,'' \emph{Satellite Navigation}, 2023.

\bibitem{chen2023satellite}
X.~Chen, R.~Luo, T.~Liu, H.~Yuan, and H.~Wu, ``Satellite navigation signal authentication in gnss: A survey on technology evolution, status, and perspective for bds,'' \emph{Remote Sensing}, 2023.

\bibitem{koisser2024orbital}
D.~Koisser, R.~Mitev, N.~Yadav, F.~Vollmer, and A.-R. Sadeghi, ``Orbital trust and privacy:$\{$SoK$\}$ on $\{$PKI$\}$ and location privacy challenges in space networks,'' in \emph{33rd USENIX Security Symposium (USENIX Security 24)}, 2024, pp. 6093--6111.

\bibitem{manulis2021cyber}
M.~Manulis, C.~P. Bridges, R.~Harrison, V.~Sekar, and A.~Davis, ``Cyber security in new space: Analysis of threats, key enabling technologies and challenges,'' \emph{International Journal of Information Security}, vol.~20, no.~3, pp. 287--311, 2021.

\bibitem{ear2023characterizing}
E.~Ear, J.~L.~C. Remy, A.~Feffer, and S.~Xu, ``Characterizing cyber attacks against space systems with missing data: Framework and case study,'' in \emph{2023 IEEE Conference on Communications and Network Security (CNS)}.\hskip 1em plus 0.5em minus 0.4em\relax IEEE, 2023, pp. 1--9.

\bibitem{pirayesh2022jamming}
H.~Pirayesh and H.~Zeng, ``Jamming attacks and anti-jamming strategies in wireless networks: A comprehensive survey,'' \emph{IEEE communications surveys \& tutorials}, vol.~24, no.~2, pp. 767--809, 2022.

\bibitem{jia2018anti}
Z.~Jia, ``Anti-jamming technology in small satellite communication,'' in \emph{Journal of Physics: Conference Series}, vol. 960, no.~1.\hskip 1em plus 0.5em minus 0.4em\relax IOP Publishing, 2018, p. 012013.

\bibitem{mirkovic2004taxonomy}
J.~Mirkovic and P.~Reiher, ``A taxonomy of ddos attack and ddos defense mechanisms,'' \emph{ACM SIGCOMM Computer Communication Review}, vol.~34, no.~2, pp. 39--53, 2004.

\bibitem{ferdous2023review}
J.~Ferdous, R.~Islam, A.~Mahboubi, and M.~Z. Islam, ``A review of state-of-the-art malware attack trends and defense mechanisms,'' \emph{IEEe Access}, vol.~11, pp. 121\,118--121\,141, 2023.

\bibitem{ATTCK}
{MITRE Corporation}, ``{MITRE ATT\&CK},'' \url{https://attack.mitre.org/}.

\bibitem{SPARTA}
{The Aerospace Corporation}, ``{SPARTA: Space Attack Research and Tactic Analysis},'' \url{https://aerospace.org/sparta}.

\bibitem{remy2025towards}
J.~L.~C. Remy and S.~Xu, ``Towards a systematic taxonomy of attacks against space infrastructures,'' \emph{arXiv preprint arXiv:2512.12829}, 2025.

\bibitem{spacethreatlandscape}
E.~Rekleitis and M.~Adamczyk, ``Space threat landscape 2025,'' 2025.

\bibitem{XuCNS2023}
E.~Ear, J.~L.~C. Remy, A.~Feffer, and S.~Xu, ``Characterizing cyber attacks against space systems with missing data: Framework and case study,'' in \emph{IEEE CNS}, 2023, pp. 1--9.

\bibitem{norgard2017electromagnetic}
J.~Norgard and G.~L. Best, ``The electromagnetic spectrum,'' in \emph{National Association of Broadcasters Engineering Handbook}.\hskip 1em plus 0.5em minus 0.4em\relax Routledge, 2017, pp. 3--10.

\bibitem{ITUfreqbands}
{ITU}, ``{ITU Recommendations V.431: Nomenclature of the frequency and wavelength bands used in telecommunications},'' \url{https://www.itu.int/rec/r-rec-v.431/en}.

\bibitem{IEEEfreqbands}
{IEEE SA}, ``{IEEE 521-2002 - Standard Letter Designations for Radar-Frequency Bands},'' \url{https://standards.ieee.org/ieee/521/768/}.

\bibitem{wertz2011space}
J.~R. Wertz, D.~F. Everett, and J.~J. Puschell, ``Space mission engineering: The new {SMAD},'' 2011.

\bibitem{remy2025quantifying}
J.~L.~C. Remy, E.~Ear, and S.~Xu, ``Quantifying and reducing system non-resilience: Methodology, metrics, and case study,'' in \emph{Cyber Resilience: Applied Perspectives}.\hskip 1em plus 0.5em minus 0.4em\relax Springer, 2025, pp. 75--103.

\bibitem{swope2024space}
C.~Swope, K.~A. Bingen, M.~Young, M.~Chang, S.~Songer, and J.~Tammelleo, ``Space threat assessment 2024,'' 2024.

\bibitem{swope2025space}
C.~Swope, K.~A. Bingen, M.~Young, and K.~Lafave, ``Space threat assessment 2025,'' 2025.

\bibitem{thiele2022investigating}
S.~Thiele and A.~C. Boley, ``Investigating the risks of debris-generating asat tests in the presence of megaconstellations,'' \emph{The Journal of the Astronautical Sciences}, vol.~69, no.~6, pp. 1797--1820, 2022.

\bibitem{kelso2007analysis}
T.~Kelso, ``Analysis of the 2007 chinese asat test and the impact of its debris on the space environment,'' in \emph{8th Advanced Maui Optical and Space Surveillance Technologies Conference, Maui, HI}, vol.~7, 2007.

\bibitem{usASATs}
{Secure World Foundation}, ``{U.S. Direct Ascent Anti-Satellite Testing},'' \url{https://cdn.prod.website-files.com/66dcc6872f6ed23bce1db235/684aeb6d1a0e4c493adcd92e_Fact%20Sheet%202025_U.S.%20Direct-Ascent%20Anti-Satellite%20Testing.pdf}.

\bibitem{weeden2020current}
B.~Weeden, ``Current and future trends in chinese counterspace capabilities,'' \emph{Proliferation Papers}, vol.~62, p.~42, 2020.

\bibitem{russianASAT}
{United States Space Command}, ``{Russian direct-ascent anti-satellite missile test creates significant, long-lasting space debris},'' \url{https://www.spacecom.mil/Newsroom/News/Article-Display/Article/2842957/russian-direct-ascent-anti-satellite-missile-test-creates-significant-long-last/}.

\bibitem{harlin2007ballistic}
W.~Harlin and D.~A. Cicci, ``Ballistic missile trajectory prediction using a state transition matrix,'' \emph{Applied mathematics and computation}, vol. 188, no.~2, pp. 1832--1847, 2007.

\bibitem{masters2013us}
J.~Masters and G.~Bruno, ``Us ballistic missile defense,'' \emph{Council on Foreign Relations}, vol.~1, 2013.

\bibitem{wright2023hypersonic}
D.~Wright and C.~L. Tracy, ``Hypersonic weapons: Vulnerability to missile defenses and comparison to marvs,'' \emph{Science \& Global Security}, vol.~31, no.~3, pp. 68--114, 2023.

\bibitem{russiaOrbitalASAT}
{Spaceflight Now}, ``{U.S. officials say Russia tested a new anti-satellite weapon},'' \url{https://spaceflightnow.com/2020/07/23/u-s-officials-say-russia-tested-a-new-anti-satellite-weapon/}.

\bibitem{NASAOrbitalCollision}
{NASA}, ``{Orbital Debris Qaurterly News - September 2025},'' \url{https://www.orbitaldebris.jsc.nasa.gov/quarterly-news/pdfs/ODQNv29i3.pdf}.

\bibitem{chineseOrbitalASAT}
{TWZ}, ``{A Chinese Satellite Just Grappled Another And Pulled It Out Of Orbit},'' \url{https://www.twz.com/44054/a-chinese-satellite-just-grappled-another-and-pulled-it-out-of-orbit}.

\bibitem{hess1964effects}
W.~N. Hess, \emph{The effects of high altitude explosions}.\hskip 1em plus 0.5em minus 0.4em\relax National Aeronautics and Space Administration, 1964.

\bibitem{snyder2025effects}
D.~Snyder, A.~Putney, E.~N. Leidy, G.~S. Hartnett, and J.~Bonomo, \emph{The Effects of High-Altitude Nuclear Explosions on Non-Military Satellites}.\hskip 1em plus 0.5em minus 0.4em\relax RAND, 2025.

\bibitem{united1976us}
U.~S.~C. on~Extension to~the Standard~Atmosphere, \emph{US standard atmosphere, 1976}.\hskip 1em plus 0.5em minus 0.4em\relax National Oceanic and Amospheric [sic] Administration, 1976.

\bibitem{earthsRadiationBelts}
{Reeves, Geoffrey D. and Delzanno, Gian Luca and Marksteiner, Quinn R}, ``{Earth’s Radiation Belts: The Hazards to Satellites and What Can Be Done to Mitigate the Risks?}'' 2020, \url{https://amostech.com/TechnicalPapers/2020/Atmospherics-Space-Weather/Reeves.pdf}.

\bibitem{zhou2014particle}
C.~Zhou and W.~Qian, ``Particle-beam weapons system,'' in \emph{2014 IEEE International Conference on Control Science and Systems Engineering}.\hskip 1em plus 0.5em minus 0.4em\relax IEEE, 2014, pp. 81--84.

\bibitem{liu2020space}
Z.~Liu, C.~Lin, and G.~Chen, ``Space attack technology overview,'' in \emph{Journal of Physics: Conference Series}, vol. 1544, no.~1.\hskip 1em plus 0.5em minus 0.4em\relax IOP Publishing, 2020, p. 012178.

\bibitem{jameel2018comprehensive}
F.~Jameel, S.~Wyne, G.~Kaddoum, and T.~Q. Duong, ``A comprehensive survey on cooperative relaying and jamming strategies for physical layer security,'' \emph{IEEE Communications Surveys \& Tutorials}, vol.~21, no.~3, pp. 2734--2771, 2018.

\bibitem{huo2017jamming}
Y.~Huo, Y.~Tian, L.~Ma, X.~Cheng, and T.~Jing, ``Jamming strategies for physical layer security,'' \emph{IEEE Wireless Communications}, vol.~25, no.~1, pp. 148--153, 2017.

\bibitem{shahid2024taxonomy}
M.~F. Shahid, K.~Mehmood, M.~Mohsin, A.~Saleem, S.~Yaqoob, and W.~Bashir, ``Taxonomy of physical layer jamming techniques and strategies for security enhancement in wireless communication: A comprehensive survey,'' 2024.

\bibitem{falco2020satellites}
G.~Falco, ``When satellites attack: Satellite-to-satellite cyber attack, defense and resilience,'' in \emph{ASCEND 2020}, 2020, p. 4014.

\bibitem{salkield2025spacejam}
E.~Salkield, S.~K{\"o}hler, S.~Birnbach, M.~Strohmeier, and I.~Martinovic, ``Spacejam: Protocol-aware jamming attacks against space communications,'' in \emph{18th ACM Conference on Security and Privacy in Wireless and Mobile Networks}, 2025, pp. 160--171.

\bibitem{hussain2014protocol}
A.~Hussain, N.~A. Saqib, U.~Qamar, M.~Zia, and H.~Mahmood, ``Protocol-aware radio frequency jamming in wi-fi and commercial wireless networks,'' \emph{Journal of communications and networks}, vol.~16, no.~4, pp. 397--406, 2014.

\bibitem{toyoshima2008ground}
M.~Toyoshima, Y.~Takayama, T.~Takahashi, K.~Suzuki, S.~Kimura, K.~Takizawa, T.~Kuri, W.~Klaus, M.~Toyoda, H.~Kunimori \emph{et~al.}, ``Ground-to-satellite laser communication experiments,'' \emph{IEEE Aerospace and Electronic Systems Magazine}, vol.~23, no.~8, pp. 10--18, 2008.

\bibitem{payne2006principles}
C.~M. Payne, \emph{Principles of naval weapon systems}.\hskip 1em plus 0.5em minus 0.4em\relax Naval Institute Press, 2006.

\bibitem{kim2023chaff}
J.-S. Kim, D.-Y. Lee, T.-H. Kim, and D.-W. Seo, ``Chaff cloud modeling and electromagnetic scattering properties estimation,'' \emph{IEEE Access}, vol.~11, pp. 58\,835--58\,849, 2023.

\bibitem{pandey2013modeling}
A.~K. Pandey, ``Modeling and simulation of chaff cloud with random orientation and distribution,'' in \emph{IEEE MTT-S International Microwave and RF Conference}.\hskip 1em plus 0.5em minus 0.4em\relax IEEE, 2013, pp. 1--4.

\bibitem{rawlins2022death}
F.~Rawlins, R.~Baker, and I.~Martinovic, ``Death by a thousand cots: Disrupting satellite communications using low earth orbit constellations,'' \emph{arXiv preprint arXiv:2204.13514}, 2022.

\bibitem{GlobalStarSimplexAttack}
{Colby Moore}, ``{Spread Spectrum Satcom Hacking: Attacking The GlobalStar Simplex Data Service},'' \url{https://blackhat.com/docs/us-15/materials/us-15-Moore-Spread-Spectrum-Satcom-Hacking-Attacking-The-GlobalStar-Simplex-Data-Service-wp.pdf}.

\bibitem{signalHijackingMaxHeadroom}
{Wikipedia}, ``{Max Headroom signal hijacking},'' \url{https://en.wikipedia.org/wiki/Max_Headroom_signal_hijacking}.

\bibitem{pentagonlaser}
D.~M. G.~R. Lamberson, ``{New World Vistas: Air and Space Power for the 21st Century, Directed Energy Volume}.''

\bibitem{tippenhauer2011requirements}
N.~O. Tippenhauer, C.~P{\"o}pper, K.~B. Rasmussen, and S.~Capkun, ``On the requirements for successful gps spoofing attacks,'' in \emph{Proceedings of the 18th ACM conference on Computer and communications security}, 2011, pp. 75--86.

\bibitem{altaweel2023gps}
A.~Altaweel, H.~Mukkath, and I.~Kamel, ``Gps spoofing attacks in fanets: A systematic literature review,'' \emph{IEEE Access}, vol.~11, pp. 55\,233--55\,280, 2023.

\bibitem{wu2020spoofing}
Z.~Wu, Y.~Zhang, Y.~Yang, C.~Liang, and R.~Liu, ``Spoofing and anti-spoofing technologies of global navigation satellite system: A survey,'' \emph{IEEE Access}, vol.~8, pp. 165\,444--165\,496, 2020.

\bibitem{longtchi2024internet}
T.~T. Longtchi, R.~M. Rodriguez, L.~Al-Shawaf, A.~Atyabi, and S.~Xu, ``Internet-based social engineering psychology, attacks, and defenses: a survey,'' \emph{Proceedings of the IEEE}, vol. 112, no.~3, pp. 210--246, 2024.

\bibitem{o2017insights}
J.~O’Leary, J.~Kimble, K.~Vanderlee, and N.~Fraser, ``Insights into iranian cyber espionage: Apt33 targets aerospace and energy sectors and has ties to destructive malware,'' \emph{Threat Research Blog}, 2017.

\bibitem{fritz2013satellite}
J.~Fritz, ``Satellite hacking: A guide for the perplexed,'' \emph{Culture Mandala}, 2013.

\bibitem{pavur2020tale}
J.~Pavur, D.~Moser, M.~Strohmeier, V.~Lenders, and I.~Martinovic, ``A tale of sea and sky on the security of maritime vsat communications,'' in \emph{2020 IEEE Symposium on Security and Privacy (SP)}, 2020.

\bibitem{willbold2023space}
J.~Willbold, M.~Schloegel, M.~V{\"o}gele, M.~Gerhardt, T.~Holz, and A.~Abbasi, ``Space odyssey: An experimental software security analysis of satellites,'' in \emph{IEEE Symposium on Security and Privacy}, 2023.

\bibitem{falco2023wannafly}
G.~Falco, R.~Thummala, and A.~Kubadia, ``Wannafly: An approach to satellite ransomware,'' in \emph{2023 IEEE 9th International Conference on Space Mission Challenges for Information Technology (SMC-IT)}.\hskip 1em plus 0.5em minus 0.4em\relax IEEE, 2023, pp. 84--93.

\bibitem{pavur2021detecting}
J.~Pavur and I.~Martinovic, ``On detecting deception in space situational awareness,'' in \emph{Proceedings of ACM AsiaCCS}, 2021, pp. 280--291.

\bibitem{pavur2019cyber}
J.~Pavur and I.~Martinovic, ``The cyber-asat: on the impact of cyber weapons in outer space,'' in \emph{2019 11th International CyCon}.\hskip 1em plus 0.5em minus 0.4em\relax IEEE, 2019.

\bibitem{smailes2023dishing}
J.~Smailes, E.~Salkield, S.~Birnbach, M.~Strohmeier, and I.~Martinovic, ``Dishing out dos: How to disable and secure the starlink user terminal,'' \emph{arXiv preprint arXiv:2303.00582}, 2023.

\bibitem{thebarge2022developing}
J.~P. Thebarge, W.~Henry, and G.~Falco, ``Developing scenarios supporting space-based ids,'' in \emph{ASCEND 2022}.\hskip 1em plus 0.5em minus 0.4em\relax American Institute of Aeronautics and Astronautics, Inc., 2022, p. 4219.

\bibitem{onen2004denial}
M.~Onen and R.~Molva, ``Denial of service prevention in satellite networks,'' in \emph{2004 IEEE International Conference on Communications (IEEE Cat. No. 04CH37577)}, vol.~7.\hskip 1em plus 0.5em minus 0.4em\relax IEEE, 2004, pp. 4387--4391.

\bibitem{usman2020mitigating}
M.~Usman, M.~Qaraqe, M.~R. Asghar, and I.~Shafique~Ansari, ``Mitigating distributed denial of service attacks in satellite networks,'' \emph{Transactions on emerging telecommunications technologies}, vol.~31, no.~6, p. e3936, 2020.

\bibitem{hitefield2018exploiting}
S.~D. Hitefield, M.~Fowler, and T.~C. Clancy, ``Exploiting buffer overflow vulnerabilities in software defined radios,'' in \emph{2018 IEEE International Conference on Internet of Things (iThings) and IEEE Green Computing and Communications (GreenCom) and IEEE Cyber, Physical and Social Computing (CPSCom) and IEEE Smart Data (SmartData)}.\hskip 1em plus 0.5em minus 0.4em\relax IEEE, 2018, pp. 1921--1927.

\bibitem{zhou2021integrated}
Y.~Zhou, Y.~Xu, and H.~Ye, ``A integrated design for structure and em stealth of microsatellite,'' in \emph{2021 International Conference on Microwave and Millimeter Wave Technology (ICMMT)}.\hskip 1em plus 0.5em minus 0.4em\relax IEEE, 2021, pp. 1--3.

\bibitem{zhao2025ultra}
C.~Zhao, M.~Jia, N.~Zhang, S.~Meng, and Y.~Tian, ``Ultra-wideband optically transparent flexible metamaterial absorber for satellite stealth,'' \emph{Scientific Reports}, vol.~15, no.~1, p. 29093, 2025.

\bibitem{sun2022stealthy}
H.~Sun and Y.~Qin, ``Stealthy configuration optimization design and rcs characteristics study of microsatellite,'' \emph{Aerospace}, vol.~9, no.~12, p. 815, 2022.

\bibitem{reiter2020spacecraft}
J.~A. Reiter, D.~B. Spencer, and R.~Linares, ``Spacecraft stealth through orbit-perturbing maneuvers using reinforcement learning,'' in \emph{AIAA Scitech 2020 Forum}, 2020, p. 0461.

\bibitem{bera2025overview}
P.~BERA, R.~Lakshmi, and H.~C. BARSHILIA, ``An overview of radar-absorbing materials and coatings for stealth application,'' \emph{SCIENCE AND CULTURE}, vol.~91, pp. 83--92, 2025.

\bibitem{wang2025review}
M.~Wang, Q.~Wang, Y.~Xiao, M.~Wang, J.~Wang, H.~Wang, and Z.~Chen, ``Review of passive shielding materials for high-energy charged particles in earth’s orbit,'' \emph{Materials}, vol.~18, no.~11, p. 2558, 2025.

\bibitem{fetter1988protecting}
S.~Fetter, \emph{Protecting Our Military Space Systems}.\hskip 1em plus 0.5em minus 0.4em\relax Center for National Policy Press, 1988.

\bibitem{yates2008systematic}
H.~Yates and M.~R. Grimaila, ``A systematic approach for securing our space assets,'' \emph{High Frontier}, 2008.

\bibitem{shabbir2018counterspace}
Z.~Shabbir and A.~Sarosh, ``Counterspace operations and nascent space powers,'' \emph{Astropolitics}, vol.~16, no.~2, pp. 119--140, 2018.

\bibitem{shapir2013lessons}
Y.~Shapir, ``Lessons from the iron dome,'' \emph{Military and Strategic Affairs}, vol.~5, no.~1, pp. 81--94, 2013.

\bibitem{goldenDome}
{Lockheed Martin}, ``{Golden Dome for America},'' \url{https://www.lockheedmartin.com/en-us/capabilities/missile-defense/golden-dome-missile-defense.html}.

\bibitem{querol2017real}
J.~Querol and A.~Camps, ``Real-time pre-correlation anti-jamming system for civilian gnss receivers,'' in \emph{Proceedings of the 30th International Technical Meeting of The Satellite Division of the Institute of Navigation (ION GNSS+ 2017)}, 2017, pp. 1267--1288.

\bibitem{vazquez2019rf}
A.~Vazquez-Castro and M.~Hayashi, ``Physical layer security for rf satellite channels in the finite-length regime,'' \emph{IEEE TIFS}, 2019.

\bibitem{schraml2021multiuser}
M.~G. Schraml, R.~T. Schwarz, and A.~Knopp, ``Multiuser mimo concept for physical layer security in multibeam satellite systems,'' \emph{IEEE Transactions on Information Forensics and Security}, 2021.

\bibitem{kalantari2015multibeam}
A.~Kalantari, G.~Zheng, Z.~Gao, Z.~Han, and B.~Ottersten, ``Secrecy analysis on network coding in bidirectional multibeam satellite communications,'' \emph{IEEE TIFS}, 2015.

\bibitem{cao2021noma}
K.~Cao, B.~Wang, H.~Ding, L.~Lv, R.~Dong, T.~Cheng, and F.~Gong, ``Improving physical layer security of uplink noma via energy harvesting jammers,'' \emph{IEEE Transactions on Information Forensics and Security}, 2021.

\bibitem{hayashi2020poisson}
M.~Hayashi and A.~Vazquez-Castro, ``Physical layer security protocol for poisson channels for passive man-in-the-middle attack,'' \emph{IEEE Transactions on Information Forensics and Security}, 2020.

\bibitem{kerns2014blueprint}
A.~J. Kerns, K.~D. Wesson, and T.~E. Humphreys, ``A blueprint for civil gps navigation message authentication,'' in \emph{2014 IEEE/ION PLANS 2014}.\hskip 1em plus 0.5em minus 0.4em\relax IEEE, 2014.

\bibitem{anderson2017chips}
J.~M. Anderson, K.~L. Carroll, N.~P. DeVilbiss, J.~T. Gillis, J.~C. Hinks, B.~W. O’Hanlon, J.~J. Rushanan, L.~Scott, and R.~A. Yazdi, ``Chips-message robust authentication (chimera) for gps civilian signals,'' in \emph{ION GNSS+ 2017}, 2017.

\bibitem{fernandez2014design}
I.~Fern{\'a}ndez-Hern{\'a}ndez, V.~Rijmen, G.~Seco-Granados, J.~Sim{\'o}n, I.~Rodr{\'\i}guez, and J.~D. Calle, ``Design drivers, solutions and robustness assessment of navigation message authentication for the galileo open service,'' in \emph{ION GNSS+ 2014}, 2014.

\bibitem{fernandez2023semi}
I.~Fernandez-Hernandez, J.~Winkel, C.~O'Driscoll, S.~Cancela, R.~Terris-Gallego, J.~A. L{\'o}pez-Salcedo, G.~Seco-Granados, A.~Dalla~Chiara, C.~Sarto, D.~Blonski \emph{et~al.}, ``Semi-assisted signal authentication for galileo: Proof of concept and results,'' \emph{IEEE Transactions on Aerospace and Electronic Systems}, 2023.

\bibitem{curran2016message}
J.~T. Curran and C.~O’Driscoll, ``Message authentication, channel coding \& anti-spoofing,'' in \emph{ION GNSS+ 2016}, 2016.

\bibitem{wu2018ecdsa}
Z.~Wu, R.~Liu, and H.~Cao, ``Ecdsa-based message authentication scheme for beidou-ii navigation satellite system,'' \emph{IEEE Transactions on Aerospace and Electronic Systems}, 2018.

\bibitem{pavur2021qpep}
J.~Pavur, M.~Strohmeier, V.~Lenders, and I.~Martinovic, ``Qpep: An actionable approach to secure and performant broadband from geostationary orbit,'' in \emph{Proc. of NDSS}, 2021.

\bibitem{huwylerqpep}
J.~Huwyler, J.~Pavur, G.~Tresoldi, and M.~Strohmeier, ``Qpep in the real world: A testbed for secure satellite communication performance,'' in \emph{Workshop on the Security of Space and Satellite Systems (SpaceSec)}, 2023.

\bibitem{yang2019anafra}
Q.~Yang, K.~Xue, J.~Xu, J.~Wang, F.~Li, and N.~Yu, ``Anfra: Anonymous and fast roaming authentication for space information network,'' \emph{IEEE TIFS}, 2019.

\bibitem{oligeri2023pastai}
G.~Oligeri, S.~Sciancalepore, S.~Raponi, and R.~D. Pietro, ``Past-ai: Physical-layer authentication of satellite transmitters via deep learning,'' \emph{IEEE TIFS}, 2023.

\bibitem{williams2020deployed}
R.~Williams, ``Deployed electromagnetic radiation deflector shield (derds) which creates a zone of minimum radiation and magnetic/plasma effects for spacecraft and extra-planetary base station protection,'' Mar.~10 2020, uS Patent 10,583,939.

\bibitem{von2009composite}
C.~J. Von~Klemperer and D.~Maharaj, ``Composite electromagnetic interference shielding materials for aerospace applications,'' \emph{Composite Structures}, vol.~91, no.~4, pp. 467--472, 2009.

\bibitem{borio2016jammer}
D.~Borio, C.~Gioia, A.~{\v{S}}tern, F.~Dimc, and G.~Baldini, ``Jammer localization: From crowdsourcing to synthetic detection,'' in \emph{Proceedings of the 29th International Technical Meeting of the Satellite Division of The Institute of Navigation (ION GNSS+ 2016)}, 2016.

\bibitem{wesson2017gnss}
K.~D. Wesson, J.~N. Gross, T.~E. Humphreys, and B.~L. Evans, ``Gnss signal authentication via power and distortion monitoring,'' \emph{IEEE Transactions on Aerospace and Electronic Systems}, 2017.

\bibitem{schmidt2020gps}
E.~Schmidt, N.~Gatsis, and D.~Akopian, ``A gps spoofing detection and classification correlator-based technique using the lasso,'' \emph{IEEE Transactions on Aerospace and Electronic Systems}, 2020.

\bibitem{humphreys2013detection}
T.~E. Humphreys, ``Detection strategy for cryptographic gnss anti-spoofing,'' \emph{IEEE Transactions on Aerospace and Electronic Systems}, 2013.

\bibitem{liu2023probabilistic}
W.~Liu and P.~Papadimitratos, ``Probabilistic detection of gnss spoofing using opportunistic information,'' in \emph{2023 IEEE/ION PLANS}.\hskip 1em plus 0.5em minus 0.4em\relax IEEE, 2023.

\bibitem{jovanovic2014multi}
A.~Jovanovic, C.~Botteron, and P.-A. Farin{\'e}, ``Multi-test detection and protection algorithm against spoofing attacks on gnss receivers,'' in \emph{2014 IEEE/ION PLANS 2014}.\hskip 1em plus 0.5em minus 0.4em\relax IEEE, 2014.

\bibitem{falletti2021performance}
E.~Falletti, G.~Falco, M.~Nicola \emph{et~al.}, ``Performance analysis of the dispersion of double differences algorithm to detect single-source gnss spoofing,'' \emph{IEEE Transactions on Aerospace and Electronic Systems}, 2021.

\bibitem{liu2021stars}
S.~Liu, X.~Cheng, H.~Yang, Y.~Shu, X.~Weng, P.~Guo, K.~C. Zeng, G.~Wang, and Y.~Yang, ``Stars can tell: a robust method to defend against $\{$GPS$\}$ spoofing attacks using off-the-shelf chipset,'' in \emph{30th USENIX Security Symposium (USENIX Security 21)}, 2021.

\bibitem{lo2018robust}
S.~Lo, Y.~H. Chen, H.~Jain, and P.~Enge, ``Robust gnss spoof detection using direction of arrival: Methods and practice,'' in \emph{ION GNSS+ 2018}, 2018.

\bibitem{ceccato2018exploiting}
S.~Ceccato, F.~Formaggio, G.~Caparra, N.~Laurenti, and S.~Tomasin, ``Exploiting side-information for resilient gnss positioning in mobile phones,'' in \emph{2018 IEEE/ION PLANS}.\hskip 1em plus 0.5em minus 0.4em\relax IEEE, 2018.

\bibitem{jansen2016multi}
K.~Jansen, N.~O. Tippenhauer, and C.~P\"{o}pper, ``Multi-receiver gps spoofing detection: Error models and realization,'' in \emph{Proceedings of ACSAC}, 2016.

\bibitem{psiaki2013gps}
M.~L. Psiaki, B.~W. O'Hanlon, J.~A. Bhatti, D.~P. Shepard, and T.~E. Humphreys, ``Gps spoofing detection via dual-receiver correlation of military signals,'' \emph{IEEE Transactions on Aerospace and Electronic Systems}, 2013.

\bibitem{jansen2018crowd}
K.~Jansen, M.~Sch{\"a}fer, D.~Moser, V.~Lenders, C.~P{\"o}pper, and J.~Schmitt, ``Crowd-gps-sec: Leveraging crowdsourcing to detect and localize gps spoofing attacks,'' in \emph{2018 IEEE SP}.\hskip 1em plus 0.5em minus 0.4em\relax IEEE, 2018.

\bibitem{clements2023dual}
Z.~Clements, T.~E. Humphreys, and P.~Ellis, ``Dual-satellite geolocation of terrestrial gnss jammers from low earth orbit,'' in \emph{2023 IEEE/ION PLANS}.\hskip 1em plus 0.5em minus 0.4em\relax IEEE, 2023.

\bibitem{lachapelle2021orbital}
D.~M. LaChapelle, L.~Narula, and T.~E. Humphreys, ``Orbital war driving: Assessing transient gps interference from leo,'' in \emph{Proceedings of the 34th ION GNSS+ 2021}, 2021.

\bibitem{spanghero2023detecting}
M.~Spanghero and P.~Papadimitratos, ``Detecting gnss misbehavior leveraging secure heterogeneous time sources,'' in \emph{2023 IEEE/ION PLANS}.\hskip 1em plus 0.5em minus 0.4em\relax IEEE, 2023.

\bibitem{spanghero2022high}
M.~Spanghero and P.~Papadimitratos, ``High-precision hardware oscillators ensemble for gnss attack detection,'' in \emph{2022 IEEE Aerospace Conference (AERO)}.\hskip 1em plus 0.5em minus 0.4em\relax IEEE, 2022, pp. 1--11.

\bibitem{spanghero2020authenticated}
M.~Spanghero, K.~Zhang, and P.~Papadimitratos, ``Authenticated time for detecting gnss attacks,'' in \emph{ION GNSS+ 2020}, 2020.

\bibitem{zhang2020protecting}
K.~Zhang, M.~Spanghero, and P.~Papadimitratos, ``Protecting gnss-based services using time offset validation,'' in \emph{2020 IEEE/ION PLANS}.\hskip 1em plus 0.5em minus 0.4em\relax IEEE, 2020.

\bibitem{boriotracking}
D.~BORIO, C.~O’DRISCOLL, and J.~FORTUNY, ``Tracking and mitigating a jamming signalwith an adaptive notch filter.''

\bibitem{sun2022anti}
Y.~Sun, F.~Chen, Z.~Lu, and F.~Wang, ``Anti-jamming method and implementation for gnss receiver based on array antenna rotation,'' \emph{Remote Sensing}, vol.~14, no.~19, p. 4774, 2022.

\bibitem{pollock1993infrared}
D.~H. Pollock, J.~S. Accetta, and D.~L. Shumaker, ``The infrared \& electro-optical systems handbook. countermeasure systems, volume 7,'' 1993.

\bibitem{trimoreau2022improving}
G.~Trimoreau, ``Improving the safe mode robustness of an earth observation satellite,'' 2022.

\bibitem{salunkhe2017analysis}
H.~S. Salunkhe, S.~Jadhav, and V.~Bhosale, ``Analysis and review of tcp syn flood attack on network with its detection and performance metrics,'' \emph{International Journal of Engineering Research \& Technology (IJERT)}, vol.~6, no.~01, pp. 2278--0181, 2017.

\bibitem{dang2019sdn}
V.~T. Dang, T.~T. Huong, N.~H. Thanh, P.~N. Nam, N.~N. Thanh, and A.~Marshall, ``Sdn-based syn proxy—a solution to enhance performance of attack mitigation under tcp syn flood,'' \emph{The Computer Journal}, vol.~62, no.~4, pp. 518--534, 2019.

\bibitem{zaoui2024comprehensive}
M.~Zaoui, B.~Yousra, S.~Yassine, M.~Yassine, and O.~Karim, ``A comprehensive taxonomy of social engineering attacks and defense mechanisms: toward effective mitigation strategies,'' \emph{IEEE access}, vol.~12, pp. 72\,224--72\,241, 2024.

\bibitem{oz2022survey}
H.~Oz, A.~Aris, A.~Levi, and A.~S. Uluagac, ``A survey on ransomware: Evolution, taxonomy, and defense solutions,'' \emph{ACM Computing Surveys (CSUR)}, vol.~54, no. 11s, pp. 1--37, 2022.

\end{thebibliography}


\end{document}